\documentclass[12pt]{article}
\usepackage{amssymb}
\def\cee{{\relax\hbox{$\inbar\kern-.3em{\rm C}$}}}

\newcommand{\be}{\begin{equation}}
\newcommand{\ee}{\end{equation}}
\newcommand{\bea}{\begin{eqnarray}}
\newcommand{\eea}{\end{eqnarray}}
\newcommand{\bml}{\begin{mathletters}}
\newcommand{\eml}{\end{mathletters}}
\newtheorem{theorem}{Theorem}

\newtheorem{definition}{Definition}

\def\darr#1{\raise1.5ex\hbox{$\leftrightarrow$}\mkern-16.5mu #1}

\def\roughly#1{\raise.3ex\hbox{$#1$\kern-.75em\lower1ex\hbox{$\sim$}}}

\def\dim{{\rm dim}}

\def\lref{\begingroup\obeylines\lr@f}
\def\lr@f#1#2{\gdef#1{\ref#1{#2}}\endgroup\unskip}

\def\and{{a^\dagger_n}}

\def\math@note#1{\gdef\@eqnlabel{LAB: #1}}
\begin{document}

\begin{flushright}
SISSA/nn/06/EP\\ hep-th/0602162n
\end{flushright}

\vspace{.1in}

\begin{center}
{\Large\bf FLUXES, BRANE CHARGES AND CHERN MORPHISMS OF HYPERBOLIC GEOMETRY}
\end{center}
\vspace{0.1in}
\begin{center}
{\large L. Bonora $^{(a)}$\footnote{ bonora@sissa.it}, A. A. Bytsenko $^{(b)}$}
\footnote{abyts@uel.br}\\
\vspace{7mm}
$^{(a)}$ {\it International School for Advanced Studies (SISSA/ISAS) \\
Via Beirut 2, 34014 Trieste, Italy  and INFN, Sezione di
Trieste} \vspace{5mm}\\
$^{(b)}$ {\it
Departamento de F\'{\i}sica, Universidade Estadual de
Londrina\\
Caixa Postal 6001, Londrina-Paran\'a, Brazil}\\
\end{center}
\vspace{0.1in}
\begin{center}
{\bf Abstract}
\end{center}
The purpose of this paper is to provide the reader with
a collection of results which can be found in the mathematical literature
and to apply them to hyperbolic spaces that may have a role
in physical theories. Specifically we apply K-theory methods for
the calculation of brane charges and
RR-fields on hyperbolic spaces (and orbifolds thereof). It is
known that by tensoring K-groups with the rationals, K-theory
can be mapped to rational cohomology by means of the Chern character
isomorphisms. The Chern character allows one to relate the
analytic Dirac index with a topological index, which can be
expressed in terms of cohomological characteristic classes. We
obtain explicit formulas for Chern character, spectral invariants,
and the index of a twisted Dirac operator associated with real
hyperbolic spaces. Some notes for a bivariant version of
topological K-theory (KK-theory) with its connection to the index
of the twisted Dirac operator and twisted cohomology of hyperbolic
spaces are given. Finally we concentrate on lower K-groups useful
for description of torsion charges.

\vfill


\section{Introduction}

K-theory methods have been applied to classify D-brane charges and
RR-fields \cite{Witten,Gukov,Sharpe}, whereby the latter have been
identified with elements of the Grothendieck K-groups
\cite{Atiyah,Karoubi,Husemoller}. In this paper we are interested
on D-branes and fluxes on hyperbolic spaces and orbifolds (see
below). Let $X = G/{\mathcal K}$ be an irreducible rank one
symmetric space of non-compact type. Thus $G$ will be a connected
non-compact simple split rank one Lie group with finite center,
and ${\mathcal K}\subset G$ will be a maximal compact subgroup.
More concretely the object of interest is the groups $G=SO_1(N,1)$
\,$(N\in {\mathbb Z}_{+})$ and ${\mathcal K}=SO(N)$. The
corresponding symmetric space of non-compact type is the real
hyperbolic space $X = {\mathbb H}^N = SO_1(N, 1)/SO(N)$ of
sectional curvature $-1$. The relevant K-theory for the latter is
the equivariant one as was shown in
\cite{Witten,Diaconescu,Garcia,Bergman}. Before going to the
specific problem considered in the paper we would like to recall
some example of string compactification with non-spherical
horizons. De Sitter, anti-de Sitter spaces and $N-$spheres
${\mathbb S}^N$ frequently appear as vacuum solutions in string
theory. These spaces (as well as $N-$dimensional real hyperbolic
spaces ${\mathbb H}^N$) naturally arise as the near-horizon of
black brane geometries. Spheres and the anti-de Sitter spaces, as
supergravity solutions, have been extensively investigated.
Research on de Sitter solutions has been limited by the fact that
they break supersymmetry and, moreover, it is hard to define a
quantum field theory on them.

Hyperbolic spaces have attracted much less attention. As an
example, let us consider a solution to the equations of motion in
eleven-dimensional supergravity
 which is provided by the Freund-Rubin ansatz for the antisymmetric
field strength. The requirement of unbroken supersymmetry, i.e. of
vanishing gravitino transformation, is equivalent to the existence
of $SO(1,6)$ and $ SO(4)$ Killing spinors $\theta$ and $\eta$,
respectively. The two-form equations admit solutions of the type
$X^7\times Y^4$, where $X^7$ and $Y^4$ are Einstein spaces of
negative and positive  curvature, respectively. But only  those
spaces that admit  Killing spinors preserve supersymmetry. The
integrability conditions for spinor equations are
$W_{\mu\nu\rho\sigma}\gamma^{\rho\sigma}\theta = 0,$\,
$W_{mnpq}\gamma^{pq}\eta = 0,$ where $W_{\mu\nu\rho\sigma}$ and
$W_{mnpq}$ are the Weyl tensors of $X^7$ and $Y^4$, respectively.
Well-known supersymmetric examples for $Y^4$ include the round
four-sphere ${\mathbb S}^4$ and its orbifolds $\Gamma\backslash
{\mathbb S}^4$, where $\Gamma$ is an appropriate discrete group
\cite{Ferrara}. For the $X^7$ space one can take the anti de
Sitter space $AdS_7$, which preserves supersymmetry as well, and
leads to the $AdS_7\!\times\!{\mathbb S}^4$ vacuum of
eleven-dimensional supergravity. There are however also solutions
involving hyperbolic spaces which are vacua of eleven-dimensional
supergravity: $ AdS_{7-N} \times {\mathbb H}^{N} \times {\mathbb
S}^4\,\, (N \geq 2)\,, AdS_{3} \times {\mathbb H}^{2} \times
{\mathbb H}^2 \times {\mathbb S}^4\,, AdS_{2}\times{\mathbb
H}^{2}\times{\mathbb H}^3\times {\mathbb S}^4. $ Hyperbolic spaces
have infinite volume with respect to the Poincar\'e metric. Thus
there are no normalizable modes for any field configurations in
hyperbolic spaces. On the other hand non-empty bulk and boundary
field theories can exist in coset spaces with topology
$\Gamma\backslash {\mathbb H}^N$.

The hyperbolic manifolds ${\mathbb H}^N$, as factors in
supergravity solutions, admit Killing spinors. However in the
space-time solutions $ AdS_{3}\times{\mathbb H}^{2}\times{\mathbb
H}^2\times{\mathbb S}^4$,\, $AdS_{2}\times{\mathbb
H}^{2}\times{\mathbb H}^3\times {\mathbb S}^4$, the factors
${\mathbb H}^{2}\times{\mathbb H}^2$,\,\, ${\mathbb
H}^{2}\times{\mathbb H}^3$ cannot leave any unbroken supersymmetry
\cite{Bytsenko22}. Although non-supersymmetric, these spaces are
an interesting setup for string compactifications, in particular
for the construction of conformal field theories.
Beside M-theory solutions, one can also have classical type II
superstring theory solutions that include hyperbolic factors.
These are, for instance, $AdS_3\times {\mathbb H}^2\times {\mathbb
S}^5$ and $ AdS_2\times {\mathbb H}^3\times {\mathbb S}^5$. Milne
universe type solutions are also found in type II: they involve
hyperbolic spaces of generic dimension \cite{BytsenkoM}. These
spaces do not have Killing spinors, therefore they do not preserve
any supersymmetry. On a general footing, supersymmetry guarantees
stability of the vacuum, but its absence does not necessarily
imply instability. In particular D-branes may well represent
stable configurations even though they are not BPS solutions.
Therefore, in the sequel, we will study the problems of D-branes
in hyperbolic spaces and relative orbifolds, abstracting from
supersymmetry considerations.

The purpose of this paper is to provide the reader with a collection of
results which can be found in the mathematical literature
and to apply them to hyperbolic spaces that may have a role
in physical theories. We refer in particular
to results obtained by cohomological and K-theory methods,
which are becoming more and more important in view of
the growing demand coming from gauge and gravity and especially
string and brane theories.

The paper is organized as follows. In Section 2 we give a simple
introduction to topological K-theory. More formal definitions and
properties can be found in Appendix, which follows \cite{Olsen}
and is designed to provide the reader with the background of
topological K-theory, especially in regards to real hyperbolic
manifolds. Section 3 is devoted to spectral functions, eta and
Chern-Simons invariants of real hyperbolic spaces. K-homology
classes describe branes: they can be associated with maps of
spin-manifolds $X$ into the spacetime background. The holonomy of
RR fields over a brane can be measured by the spectral eta
invariant of a vector bundle restricted to $X$. We consider
spectral functions of hyperbolic geometry in Section 3.1. Using
the standard definition for the Dirac bundle we review results on
the spectral Selberg type zeta functions (or Shintani functions)
and the twisted holomorphic eta function of Atiyah-Patodi-Singer.
In Section 3.2 we describe the explicit formula for
$U(n)-$Chern-Simons invariant of hyperbolic three-spaces. If the
three-manifold is a homology three-sphere (like in our case),
every $U(n)-$representation of the fundamental group is a
$SU(n)-$representation. A connection between the index theory of
Dirac operators and topological K-theory is presented in Section
4. Dirac operators are examples of pseudo-differential elliptic
operators, which are Fredholm when viewed as operators on a
Hilbert space. The indexes of the twisted Dirac operators acting
on spaces with topology $X=SO_1(2n,1)/SO(2n)$ are given in Section
4.1. The Kasparov's KK-pairings and the concept of K-amenable
groups are considered in Section 5.1. In Section 5.2 we give some
notes on K-groups of symmetric spaces, which are useful for the
description of D-brane torsion charges. Finally in Section 6 we
concentrate on the lower algebraic K-groups of a real compact
oriented hyperbolic three-manifold.

\section{Aspects of topological K-theory}

In this section we would like to review some general ideas and
collect some formulas of K-theory in type IIA and type IIB
superstring theories, with a view to applying them to a hyperbolic
background. Following \cite{Olsen} we have collected a few very
basic definitions and properties concerning these groups in
Appendix. Here we mostly stick to a rather informal presentation.

It is well-known that type IIB brane charges in a space-time
manifold $X$ are well described as elements of $K(X)$ and type IIA
brane charges as elements of $K^{-1}(X)$, while RR-fluxes are
classified by $K^{-1}(X)$ and $K(X)$, respectively. Let us start
from the simplest example: a type IIB theory in which $X$ is
non-compact and is the product of the brane world-volume ${\mathbb
R}^{p+1}$ and the transverse space ${\mathbb H}^N$. In this
situation the charge is determined by the transverse
space\footnote{This comes from the equation (Bianchi identity) $dF_p
=\delta(W)$ characteristic of a $p-$brane with world-volume $W$:
$\delta(W)$ is $(p+1)-$delta-function-form which is the Poincar\'e
dual of the submanifold $W$. This form is `transverse' to $W$ and
determines a cohomology class, which gets lifted to a K-theory
class in the K-theoretic framework. By transverse directions we
refer to the fibres of the normal bundle of $W$ in $X$, the
relevant K-theory being the one with compact support along the
fibre.}.

The charge in the simplest cases is just an integral over a form
density. At times the form class itself is referred to as the
charge and we will also stick to this small abuse of language. In
general however cohomology is not enough,
\cite{Moore,Witten,Moore00}: cohomology has to be lifted to
K-theory in order to describe brane charges appropriately. A
cohomology class is replaced by a K-theory class. So charges are
represented by K-theory classes. In the simple example we are
considering the relevant space is the transverse space ${\mathbb
H}^N$. It is homeomorphic to ${\mathbb R}^N$, therefore its
K-theory is going to be the same as for the latter. If ${\mathbb
H}^N$ is the transverse space (fibre) to a (non-compact) brane, by
definition we set $K({\mathbb H}^N) \equiv\widetilde K({\mathbb
S}^N)$ where $\widetilde K$ is the reduced K-theory. Since
$\widetilde K({\mathbb S}^N)=\delta_{N,0}\,{\mathbb Z}$ for $N\,\,
{\rm mod}\, 2$, we get for $\ell \in {\mathbb Z}_+$ the following
result
\begin{equation}
K({\mathbb H}^N) =\left\{\matrix{\!\!\!\!\!\!\!\!\!
{\mathbb Z},\quad N = 2\ell\,,\cr
0, \quad N = 2\ell+1\,.\cr}\right.
\label{KHn}
\end{equation}
Reduced K-groups are formally defined in Appendix: $K(X)$ being
the Grothendieck group defined by copies $(E,F)$ of vector bundles
over X, the reduced K-group is defined by copies of vector bundles
of the same rank. This is physically motivated by the fact that,
in the absence of D-branes in the vacuum, we require any two
bundles $E$ and $F$ to be isomorphic `at infinity', \cite{Witten}.
In turn $\tilde K(X)$ is isomorphic to $K_c(X)$, the K-group with
compact support.

For completeness we write down the only other distinct K-groups
(due to periodicity, see Appendix)
\begin{equation}
K^1({\mathbb H}^N )\equiv \widetilde K^1({\mathbb S}^N)
=K^1({\mathbb S}^N)=
K({\mathbb S}^{N+1})= {\mathbb Z}\oplus \widetilde K({\mathbb S}^{N+1})\,
\mbox{.}
\label{K1Hn}
\end{equation}
These groups are relevant to the classification of RR-fluxes.

As pointed out in the introduction, supergravity solutions which
are noncompact in the internal directions are not very appealing.
Therefore the previous examples have only a pedagogical
motivation. In the sequel we wish to discuss more interesting
cases involving compact hyperbolic spaces obtained from
non-compact ones by acting some discrete group $\Gamma$. In this
case the simplified treatment given above is not adequate. First
of all the relevant K-theory is the equivariant one, $K(X_\Gamma)$
(see Appendix); moreover the `transverse directions' must be
replaced by the consideration of the normal bundle to the brane
world-volume. In general the K-groups have a free part and a
torsion part. By tensoring the K-groups with the rationals it is
possible to map the K-theory to rational cohomology and obtain
very interesting formulas. In this passage a crucial role is
played by the Chern character isomorphism. Let us review it.

{\it The Chern character}. One of the features of the topological
K-theory which makes it so useful in a variety of applications is
the existence of the Chern character isomorphism.
To set the necessary background for
our discussion, we shall briefly review the construction of the
Chern character in the topological K-theory. Let ${\mathbb E}$ be
a complex vector bundle over a compact ${\it topological}$ space
${X}$. There are  special cohomology classes of ${\mathbb E}$ ,
the Chern characteristic classes: $c_j({\mathbb E}) \in
H^{2j}({X}, {\mathbb Z})$. Their basic properties are the
following ones:

\vspace{0.5cm}

 {({\rm {\bf i}})\,\,\,\,\,\,  $c_0({\mathbb E}) = 1 \in H^0({X},
{\mathbb Z})$;}

 {({\rm {\bf ii}})\,\,\,\, For all $n\geq 0 , c_n({\mathbb E}\oplus {\mathbb F})
= \sum_{i+j= n } c_i({\mathbb E}) \cup c_j({\mathbb F})$;}

 {({\rm
{\bf iii}})\,\, If $f: {Y}\rightarrow {X}$ is a continuous
map, then $c_n(f^*{\mathbb E}) = f^*c_n({\mathbb E})$.}

\vspace{.5cm}

In fact the axiom $({\rm {\bf ii}})$ is the Whitney sum formula. It
implies that the total Chern class $c({\mathbb E}) =
\sum_{j >0} c_j({\mathbb E})\in H^\#({X}, {\mathbb Z})$
depends only on the class of the ${\mathbb E}$ bundle
in $K^0({X})$.
Therefore $c_j$ extend to functions
$c_j: K^0({X}) \rightarrow H^{2j}({X}, {\mathbb Z})$.
We can thus, for any complex vector bundle ${\mathbb E}$,
define the Chern
character $ch$ with the properties:

\vspace{.5cm}

{({\rm {\bf iv}})\,\,\,\,\,\, ${\rm ch}_0({\mathbb E})$ equals the
rank $({\mathbb E})$ of ${\mathbb E}$, where rank$({\mathbb E})
\in H^0({X}, {\mathbb Z}) \simeq {\mathbb Z}$;}

{ ({\rm {\bf v}})\,\,\,\,\,\, ${\rm ch} ({\mathbb E} \oplus
{\mathbb F}) = {\rm ch}({\mathbb E}) + {\rm ch}({\mathbb F})$;}

 {
({\rm {\bf vi}})\,\,\,\,\,\,
${\rm ch}(f^*({\mathbb E}))= f^*({\rm ch}({\mathbb E}))$
for a continuous map $f: {Y}\rightarrow {X}$. }

\vspace{0.5cm}

The Chern character can be viewed as a natural transformation of
functors $K^\# (\bullet) \rightarrow H^\#(\bullet, {\mathbb Q})$.
Eliminating torsion by using rational K-theory, we get the
following result. Let $X$ be a finite $CW-$complex, then the Chern
character $ {\rm ch} : K^\# (X)\otimes {\mathbb Q} \rightarrow
H^\# (X;{\mathbb Q}) $ is a natural isomorphism between rational
K-theory and rational cohomology. The Chern characters ${\rm
ch}^\#$ (cohomology) and ${\rm ch}_\#$ (homology) preserve the
``cap'' product $\cap$. It means that for every {\it topological
space} $X$ there is a ${\mathbb Z}_2-$degree preserving
commutative sequence \cite{Switzer,Reis}:
\begin{eqnarray}
&&
K^\#(X)\otimes K_\#(X)\stackrel{\bigcap}{\longrightarrow} K_\#(X)
\stackrel{{\rm ch}_\#}{\longrightarrow} H_\#(X, {\mathbb Q})
\nonumber \\
&&
H_\#(X, {\mathbb Q})
\stackrel{\bigcap}{\longleftarrow}H^\#(X, {\mathbb Q})\otimes
H_\#(X, {\mathbb Q})\stackrel{{\rm ch}^\#\bigotimes {\rm ch}_\#}
{\longleftarrow}K^\#(X)\otimes K_\#(X)
\end{eqnarray}

The Chern isomorphism allows us to arrive at a very nice formula
for rational D-brane charges \cite{Cheung,Moore}. To this end we
need two more tools, the Gysin homomorphism and the Thom
isomorphism.

{\it Gysin and Thom maps}.
Before discussing the brane charge formula, we begin with some
conventions which apply throughout. Let $X$ be an oriented manifold,
and let $H^\#(X)$ be the cohomology ring of $X$. The Poincar\'e
duality (a well-known result in differential topology) gives a
canonical isomorphism
\begin{equation}
{\mathfrak d}_X:\,\,\,
H^\ell(X)\stackrel{\approx}{\longrightarrow} H_{n-\ell}(X)\,,
\,\,\,\,\, {\rm for}\,\,\,\, {\rm all}\,\,\,\, \ell =0,1,\ldots,
\,\,\, n={\rm dim}\,X\,.
\end{equation}
Let $f:Y\rightarrow X$ be a continuous map from $Y$ to $X$ and $m=
{\rm dim}\,Y$. For all $\ell\geq m-n$ there is a linear map,
called the Gysin homomorphism: $
f_{!}:\,H^{\ell}(Y)\longrightarrow H^{\ell-(m-n)}(X)$, which is
defined such that the diagram
\begin{eqnarray}
\matrix{H^\ell(Y)& \stackrel{{\mathfrak d}_Y }{\longrightarrow}
&H_{m-\ell}(Y)\cr {f_!}\downarrow& & \downarrow f_*\cr
 H^{\ell-(m-n)}(X)&\stackrel{{\mathfrak d}_X^{-1}}{\longleftarrow}&
H_{m-\ell}(X)}\label{commdiagr}
\end{eqnarray}
is commutative. Thus, $f_{!}={\mathfrak d}_X^{-1}\,f_*\,{\mathfrak
d}_Y$, where $f_*$ is the natural push-forward map acting on
homology. As an example of that construction, let us assume that $Y$
is an oriented vector bundle $E$ over $X$, of fiber dimension
$\ell$. The canonical projection map, $\pi:E\rightarrow X$, and the
inclusion $\iota: X\rightarrow E$ of the zero section induce maps on
the homology with $\pi_*\iota_*={\rm Id}$. For all $j$, we have the
following isomorphisms:
$$ \pi_{!}:\, H^{j+\ell}(E)
\stackrel{\approx}{\longrightarrow} H^j(X)\,,\,\,\,\,\,
\iota_{!}:\, H^j(X) \stackrel{\approx}{\longrightarrow} H^{j+\ell}(E)\,.
$$
$\pi_{!}$ is the Gysin map; it can be associated with integration
over the
fibers of $E\rightarrow X$. We have $\pi_{!}\iota_{!}={\rm Id}$, so
that $\pi_{!}=(\iota_{!})^{-1}$. The map $i_{!}$ is called the Thom
isomorphism of the oriented vector bundle $E$. The particular
example $j=0$ is an important case of the Thom isomorphism. For
$j=0$, a map $H^0(X)\rightarrow H^{\ell}(E)$ and the image of $1\in
H^0(X)$ determine a cohomology class $\iota_!(1)\in H^{\ell}(E),$
which is called the Thom class of $E$.

After this preliminary review, let us consider a $U(n)$ gauge
bundle ${\mathbb E}$ on the brane. It has been shown,
\cite{Moore}, that, using the Gysin and Thom maps, one can find an
explicit expression for rational brane charges. As an element of
$H^{*}(X)$, the RR charge (class) associated with a D-brane
wrapping a supersymmetric cycle
in spacetime $f:Y\hookrightarrow X$ with
Chan-Paton bundle ${\mathbb E}\rightarrow Y$, is given by
\begin{equation}
Q={\rm ch}(f_!{\mathbb E})\wedge [\widehat{A}(TX)]^{1/2}\,
\mbox{.} \label{Qfinal}
\end{equation}
One can obtain charges by integrating the RHS over the appropriate cycles.
As explained above the map
\begin{equation}
{\rm ch}: \, K(X)\otimes_{\mathbb Z}{\mathbb Q}\longrightarrow
H^{\rm even}(X,{\mathbb Q})
\equiv\bigoplus_{n\geq 0}H^{2n}(X,{\mathbb Q})
\end{equation}
is an isomorphism, and it can be extended to a ring isomorphism
\cite{Atiyah62}, $ {\rm ch}: \, K^{*}(X)\otimes_{\mathbb Z}{\mathbb
Q} \stackrel {\approx}{\longrightarrow} H^{*}(X,{\mathbb Q}), $
which maps $K^{-1}(X)\otimes_{\mathbb Z}{\mathbb Q}$ onto $H^{\rm
odd}(X,{\mathbb Q})$.
Note that the rational cohomology ring $H^{\rm even}(X, {\mathbb
Q})$ has a natural inner product, while the pairing $K(X)$,
associated with the cohomology ring  $K(X)\otimes_{\mathbb
Z}{\mathbb Q}$, is given by the index of the Dirac operator. The
result (\ref{Qfinal}) is in complete agreement with the fact that
the D-brane charge is given by $f_![{\mathbb E}]\in K(X)$, and it
gives an explicit formula for the brane charges in terms of the
Chern character isomorphism on K-theory.

So far we have been dealing with rational K-theory. The torsion
part is more difficult to deal with. However there are instances
in which concrete results can be obtained. In the next section we
will be exploiting the following remark.

For a finite-dimensional ${\it smooth}$ ${\it manifold}$ $X$ the
RR-phase admits a description in terms of the exact sequence
\cite{Boer}:
$$
0 \rightarrow H^{\rm odd}(X,{\mathbb R})/[K^1(X)/{\rm Tor}]
\rightarrow K^1(X,U(1)) \rightarrow {\rm Tor} K^0(X)
\rightarrow 0
$$
The component group is given by the torsion classes in $K^0(X)$,
while the trivial component consists of the torus $H^{\rm
odd}(X,{\mathbb R})/(K^1(X)/{\rm Tor})$. If $H^{\rm
odd}(X,{\mathbb R})$ vanishes, the group of fluxes is simply $
K^{1}(X,U(1)) \cong {\rm Tor} K^0(X), $ and the corresponding RR-
fluxes can be represented by flat vector bundles. Since $K^0(X)$
measures charges in type IIB string, this amounts to saying that
the flat RR-fluxes $K^{1}(X,U(1))$ coincide with the torsion
charges ${\rm Tor} K^0(X)$.

\section{Chern classes and Chern-Simons invariants}

In the string theory path integral a torsion RR-flux prduces an
additional phase factor to a D-brane. A brane can be represented
by a K-homology class, that is a map of
 an additional Chan-Paton vector bundle into the spacetime background
\cite{Boer}. The holonomy of the RR-fields over this brane can be
associated with the spectral eta invariant of a vector bundle
restricted to $X$. For $X$ a closed oriented hyperbolic manifold
and $\chi$ an orthogonal
 representation of $\pi_1X$, we suppose
that $\chi$ is acyclic (it means that the vector space of twisted
cohomology classes vanishes). Then the eta invariant can be computed
from the twisted cochain complex (see for more detail the next section).
 In the sequel we consider the spectral functions and explicitly
calculate the eta and Chern-Simons invariants of hyperbolic
spaces, related to the holonomy of RR-fields over branes.

\subsection{Complex spectral functions of hyperbolic geometry}

Let $X_{\Gamma}= \Gamma\backslash G/{\mathcal K}$ be a real
compact hyperbolic manifold. The fundamental group of $X_{\Gamma}$
acts by covering transformations on $X$ and gives rise to a
discrete, co-compact subgroup $\Gamma \subset G$ ($\Gamma$ is
called co-compact if it contains only hyperbolic elements). Let
${\mathfrak D}$ denote a generalized Dirac operator associated to
a locally homogeneous Clifford bundle over a compact oriented odd
dimensional locally symmetric space $X_{\Gamma}$, whose simply
connected cover $\widetilde{X}$ is a symmetric space of noncompact
type. The fixed point set of the geodesic flow, acting on the unit
sphere bundle $T^1X_{\Gamma}$, is a disjoint union of submanifolds
$X_{\gamma}$. These submanifolds are parametrized by the
nontrivial conjugacy classes $[\gamma] \neq 1$ in $\Gamma=\pi_1X$.
By ${\mathcal  E}_1(\Gamma)$ we denote the set of those conjugacy
classes $[\gamma]$ for which $X_{\gamma}$ has the property that
the Euclidean de Rham factor of $\widetilde{X}_{\gamma}$ is
one-dimensional; it means that for $[\gamma] \in {\mathcal
E}_1(\Gamma)$, $\widetilde{X}_{\gamma}\cong {\mathbb R}\times
\widetilde{X}_{\gamma}^\prime$ and the lines ${\mathbb R}\times
\{x^\prime\}$, $x^\prime \in \widetilde{X}_{\gamma}^\prime$ are
the axes of $\gamma$. Projected down to $X_{\gamma}$, they become
closed geodesic $c_{\gamma}$, which foliate $X_{\gamma}$. The
`center' bundle $C\widehat{X}_{\gamma}$ over the space of leaves
$\widehat{X}_{\gamma}$, which in fact turn out to be an orbifold,
is determined by the eigenvalues of absolute value 1 of the linear
Poincar\'{e} map $P(\gamma)$ \footnote{ At each point of
$X_{\gamma}$ one can consider the restriction $P_h(\gamma)$ of the
linear Poincar\'{e} map $P(\gamma)= d\Phi_1$, i.e. the
differential $d\Phi_{t=1}$ of the geodesic flow $\Phi_t$ at
$(c_{\gamma}, \dot{c}_{\gamma})\in TX_{\gamma}$, to the direction
normal to the $\Phi_t$.}. The parallel translation around closed
geodesics $c_{\gamma}$ gives rise to an orthogonal transformation
${\widehat \tau}_{\gamma}$ of $C\widehat{X}_{\gamma}$;
$T\widehat{X}_{\gamma} \subset C\widehat{X}_{\gamma}$ and we let
$N\widehat{X}_{\gamma}$ denote the orthogonal component of
$T\widehat{X}_{\gamma}$ in $C\widehat{X}_{\gamma}$. The tangent
bundle $T\widehat{X}_{\gamma}$ corresponds to the eigenvalue 1 of
${\widehat \tau}_{\gamma}$; $N\widehat{X}_{\gamma}$ can be
decomposed according to the other eigenvalues --1, $\exp(\pm
\sqrt{-1}\theta)$\, $(0<\theta<\pi)$. For more details we refer
the reader to the paper \cite{Moscovici1}.

The restriction to $X_{\gamma}$ of the exterior bundle $\Lambda
T_{\mathbb C}X_{\Gamma}$ can be pushed down to a vector bundle
${\widehat \Lambda}_{\gamma}$ over $\widehat{X}_{\gamma}$ which
splits into a subbundle ${\widehat {\Lambda}}_{\gamma}^{\pm}$
corresponding to the eigenvalue $\pm \sqrt{-1}$ of the symbol
${\mathfrak D}$. The Clifford multiplication induces a
homomorphism ${\widehat \sigma}_{\gamma}^{\mathfrak D}$ of vector
bundles associated with projection map of ${\widehat X}_{\gamma}$.
We obtain a ${\widehat \tau}_{\gamma}-$equivariant complex
${\widehat \sigma}_{\gamma}^{\mathfrak D}: {\widehat
\Lambda}_{\gamma}^{+} \rightarrow {\widehat \Lambda}_{\gamma}^{-}$
over $T\widehat{X}_{\gamma}$ and a class $[{\widehat
\sigma}_{\gamma}^{\mathfrak D}] \in K_{{\widehat \tau}_{\gamma}}^0
(T\widehat{X}_{\gamma})$, the ${\widehat
\tau}_{\gamma}-$equivariant K-theory group of
$T\widehat{X}_{\gamma}$. The cohomology class can be formed as in
\cite{Atiyah} (Section 3),
\begin{equation}
{\rm ch}({\widehat \sigma}_{\gamma}^{\mathfrak D} ({\widehat
\tau}_{\gamma})) \in H^{even}(T\widehat{X}_{\gamma}; {\mathbb C})
\mbox{.}
\end{equation}
\begin{theorem} (H. Moscovici and R. J. Stanton \cite{Moscovici1})
\,\,\, The following Selberg type function $Z(s, {\mathfrak D})$
can be defined, initially for $\Re(s^2)\gg 0$, by the formula
\begin{equation}
{\rm log}Z(s,{\mathfrak D})\stackrel{def}{=}
\sum_{[\gamma]\in {\mathcal  E}_1(\Gamma)}
(-1)^q\frac{L(\gamma,{\mathfrak D})}
{|{\rm det}(I-P_h(\gamma))|^{1/2}}\frac{e^{-s\ell(\gamma)}}{m(\gamma)}
\mbox{,}
\end{equation}
where $q=(1/2){\rm dim}(N\widehat{X}_{\gamma})$ is an integer
independent of $\gamma$, $[\gamma]$ runs over the nontrivial
conjugacy classes in $\Gamma$, $\ell (\gamma)$ is the length of
the closed geodesic $c_{\gamma}$ in the free homotopy class
corresponding to $[\gamma]$, $m(\gamma)$ is the multiplicity of
$c_{\gamma}$. The Lefschetz number $L(\gamma,{\mathfrak D})$ is
given by (see, for example, \cite{Hotta} or \cite{Moscovici1}, Eq.
(5.5)):
\begin{equation}
L(\gamma,{\mathfrak D}) = \left\{\frac{{\rm ch}({\widehat
\sigma}_{\gamma}^{\mathfrak D} ({\widehat \tau}_{\gamma}))
{\mathfrak R}(N {\widehat X}_{\gamma}(-1))
\prod_{0<\theta<\pi}{\mathfrak S}^{\theta} (N{\widehat
X}_{\gamma}(\theta)){\mathfrak T}({\widehat X}_{\gamma})} {{\rm
det}(I-{\widehat \tau}_{\gamma}|N{\widehat X}_{\gamma})} \right\}
[T{\widehat X}_{\gamma}] \mbox{.} \label{Lef}
\end{equation}
The Lefschetz formula (\ref{Lef}) has been given using the stable
characteristic classes ${\mathfrak R}, {\mathfrak S}^{\theta}$ and
${\mathfrak T}$ defined in \cite{Atiyah11} (Theorem 3.9).
\end{theorem}

Furthermore ${\rm log}Z(s,{\mathfrak D})$ has a meromorphic
continuation to ${\mathbb C}$ given by the identity
\begin{equation}
{\rm log}Z(s,{\mathfrak D})={\rm log}{\rm det}^{\prime}
\left(\frac{{\mathfrak D}-\sqrt{-1}s}{{\mathfrak D}+\sqrt{-1}s}\right)
+\sqrt{-1}\pi\eta(s,{\mathfrak D})
\mbox{,}
\end{equation}
where $s\in \sqrt{-1}{\rm Spec}^{\prime}({\mathfrak D})$\,\,\,
$({\rm Spec}^{\prime}\equiv{\rm Spec}({\mathfrak D})-\{0\})$, and $Z(s,{\mathfrak D})$
satisfies the functional equation
\begin{equation}
Z(s,{\mathfrak D})Z(-s,{\mathfrak D})=
e^{2\pi\sqrt{-1}\eta(s,{\mathfrak D})}
\mbox{.}
\label{Z1}
\end{equation}

{\it Generalized Dirac operator}. From now on we restrict
ourselves to bundles that satisfy a local homogeneity condition: a
vector bundle $E$ over $X_{\Gamma}$ is ${\mathcal G}-$locally
homogeneous, for some Lie group ${\mathcal G}$, if there is a
smooth action of ${\mathcal G}$ on $E$ which is linear on the
fibers and covers the action of ${\mathcal G}$ on $\widetilde{X}$.
Standard constructions from linear algebra applied to any
${\mathcal  G}-$locally homogeneous ${E}$ give in a natural way
corresponding ${\mathcal  G}-$locally homogeneous vector bundles.
In particular, all bundles $TX_{\Gamma}, {\mathbb
C}\ell(X_{\Gamma}), {\rm End}\,{E}\simeq {E}^{*}\otimes {E}$ are
${\mathcal  G}-$locally homogeneous \cite{Moscovici1}. We suppose
that all constructions associated with ${\mathcal G}-$locally
homogeneous bundles are ${\mathcal  G}-$equivariant.
Let ${\mathfrak D}$ denote a generalized Dirac operator
associated to a locally
homogeneous bundle $E$ over $X_{\Gamma}$. We shall
require ${\mathcal  G}-$equivariance for the natural connection
$\nabla$ so that the corresponding Dirac operator is
${\mathcal G}-$invariant.

Suppose now that $\chi: \Gamma\rightarrow U(F)$ be a unitary
representation of $\Gamma$ in $F$. The Hermitian vector bundle
${\mathbb F}= X\times_{\Gamma}F$ over $X_{\Gamma}$
inherits a flat connection from the trivial connection on
$X\times F$. If ${\mathfrak D}:
C^{\infty}(X,V)\rightarrow C^{\infty}(X,V)$ is a differential
operator acting on the sections of the vector bundle $V$, then
${\mathfrak D}$ extends canonically to a differential operator
${\mathfrak D}_{\chi}: C^{\infty}(X,V\otimes{\mathbb
F})\rightarrow C^{\infty}(X,V\otimes{\mathbb F})$, uniquely
characterized by the property that ${\mathfrak D}_{\chi}$ is
locally isomorphic to ${\mathfrak D}\otimes...\otimes {\mathfrak
D}$\,\,\, (${\rm dim}\,F$ times). We specialize to the case of
locally homogeneous Dirac operators ${\mathfrak D}:
C^{\infty}(X_{\Gamma},{\mathbb E})\rightarrow C^{\infty}
(X_{\Gamma},{\mathbb E})$ in order to construct a generalized
operator ${\mathfrak D}_{\chi}$, acting on spinors with
coefficients in $\chi$ (see for detail \cite{Moscovici1}). One can
repeat the arguments of the previous discussion to construct a
twisted zeta function $Z(s,{\mathfrak D}_{\chi})$.

\begin{theorem} (H. Moscovici and J. J. Stanton \cite{Moscovici1}).
\,\,\,There exists a zeta function  $Z(s,{\mathfrak D}_{\chi})$,
meromorphic on ${\mathbb C}$,
given for $\Re(s^2)\gg 0$ by the formula
\begin{equation}
{\rm log}Z(s,{\mathfrak D}_{\chi})\stackrel{def}{=}
\sum_{[\gamma]\in {\mathcal  E}_1(\Gamma)}
(-1)^q{\rm Tr}\chi(\gamma)\frac{L(\gamma,{\mathfrak D})}
{|{\rm det}(I-P_h(\gamma))|^{1/2}}\frac{e^{-s\ell(\gamma)}}{m(\gamma)}
\mbox{;}
\end{equation}
moreover, one has
\begin{equation}
{\rm log}Z(0,{\mathfrak D}_{\chi})=
\sqrt{-1}\pi\eta (0,{\mathfrak D}_{\chi})
\mbox{.}
\label{Z2}
\end{equation}
\end{theorem}

\subsection{$U(n)-$gauge bundles and the Chern-Simons invariants}

The Chern character allows one to map the analytical Dirac index
in terms of K-theory classes into a topological index which can be
expressed in terms of cohomological characteristic classes. This
results in a connection between the Chern-Simons action
and the celebrated Atiyah-Singer
index theorem. The goal of this section is to present explicit
formulas for the Chern classes and $U(n)-$Chern-Simons invariant
of an irreducible flat connection on real compact hyperbolic
three-manifolds.

The Chern-Simons functional $CS$ as a function on the space of
connections on a trivial principal bundle over a compact oriented
three-manifold $X_{\Gamma}$ is given by
$$ CS(A)=
\frac 1{8\pi^2}\int_{X_{\Gamma}}\mbox{Tr} \left(A\wedge dA+ \frac 23\,A
\wedge A\wedge A\right)\,. $$ Let ${\mathcal P}=X_{\Gamma}\otimes
{\mathfrak G}$ be a principal bundle over $X_{\Gamma}$ with the
gauge group ${\mathfrak G}=U(n)$ and let ${\mathfrak
A}_{X_{\Gamma}}=\Omega^1(X_{\Gamma};{\mathfrak g})$ be the space
of all connections on ${\mathcal P}$; this space is an affine
space of one-forms on $X_{\Gamma}$ with values in the Lie algebra
${\mathfrak g}$ of ${\mathfrak G}$. The value of the function
$CS(A)$ on the space of connections ${\mathfrak A}_{X_{\Gamma}}$
 can be regarded as a topological invariant of a pair
$(X_{\Gamma},\chi)$,
where $\chi$ is an orthogonal representation of the
fundamental group $\Gamma$.
Let $U_{X_{\Gamma}}=\{A\in{\mathfrak A}_{X_{\Gamma}}|F_A=
dA+A\wedge A=0\}$
be the space of flat connections on ${\mathcal P}$.

A well-known formula related to the $CS$ integrand is: $ d{\rm
Tr}(A\wedge dA+ (2/3)A\wedge A\wedge A) = {\rm
Tr}\left(F_{A}\wedge F_{A}\right)\,. $ This formula permits
another approach to the Chern-Simons invariant. Indeed, let $M$ be
an oriented four-manifold with boundary $\partial M=X_{\Gamma}$.
One can extend ${\mathcal P}$ to a trivial ${\mathfrak G}-$bundle
over $M$; then Stokes' theorem gives
\begin{equation}
CS({\widetilde A})=
\frac{1}{8\pi^2}\int_{M}\mbox{Tr}\left(F_{{\widetilde A}}
\wedge F_{{\widetilde   A}}\right)
\mbox{,}
\end{equation}
where ${\widetilde A}$ is any extension of $A$ over $M$. This can
be viewed as a generalization of the Chern-Simons invariant to the
case in which $\mathcal P$ is a non-trivial $U(n)-$bundle over
$X_{\Gamma}$. Suppose that $\chi$ is any one-dimensional
representation of a $\Gamma$ factor through a representation of
$H^1(X;{\mathbb Z})$. It can be shown that for a unitary
representation $\chi: \Gamma\rightarrow U(n)$, the corresponding
flat vector bundle ${\mathbb E}_{\chi}$ is topologically trivial
$({\mathbb E}_{\chi}\cong X\otimes {\mathbb C}^n)$ if and only if
\begin{equation}
{\rm det}\,\chi|_{{\rm Tor}^1}: {\rm Tor}^1\rightarrow U(1)
\end{equation}
is the trivial representation. Here ${\rm Tor}^1$ is the torsion
part of $H^1(M;{\mathbb Z})$ and ${\rm det}\,\chi$ is a
one-dimensional representation of $\Gamma$ defined by ${\rm
det}\,\chi(\gamma) :={\rm det}\,(\chi(\gamma))$, for $\gamma\in
\Gamma$. If ${\widetilde  A}_{\chi}$ is an extension of a flat
connection $A_{\chi}$ corresponding to $\chi$, the second Chern
character $ {\rm ch}_2({\mathbb {\widetilde E}}_{\chi})\, (=
-(1/8\pi^2){\rm Tr}(F_{{\widetilde A}_{\chi}}\wedge F_{{\widetilde
A}_{\chi}})) $ of ${\mathbb {\widetilde   E}}_{\chi}$ can be
expressed in terms of the first and second Chern classes: $ {\rm
ch}_2({\mathbb {\widetilde   E}}_{\chi})= (1/2)c_1({\mathbb
{\widetilde E}}_{\chi})^2- c_2({\mathbb {\widetilde   E}}_{\chi})
\mbox{.} $ The Chern character and the $\widehat A-$genus, the
usual polynomial related to Riemannian curvature $\Omega^M$,
${\widehat  A}(\Omega^M)=\sqrt{{\rm det}(\frac{\Omega^M/4\pi}
{{\rm sinh}\Omega^M/4\pi})}$, are given by
\begin{eqnarray}
{\rm ch}({\mathbb {\widetilde   E}}_{\chi}) & = & {\rm rank}\,
{\mathbb {\widetilde   E}}_{\chi}+
c_1({\mathbb {\widetilde   E}}_{\chi})
+{\rm ch}_2({\mathbb {\widetilde   E}}_{\chi})
={\rm dim}\,\chi +c_1({\mathbb {\widetilde   E}}_{\chi})+
{\rm ch}_2({\mathbb {\widetilde   E}}_{\chi})
\mbox{,}
\nonumber \\
{\widehat  A}(\Omega^M) & = & 1-\frac{1}{24}p_1(\Omega^M)
\mbox{.}
\end{eqnarray}
Here $p_1(\Omega^M)$ is the first Pontrjagin class, $\Omega^M$
is the Riemannian curvature of a four-manifold $M$ which has a boundary
$\partial M = X_{\Gamma}$. Thus we have
\begin{eqnarray}
{\rm ch}({\mathbb {\widetilde E}}_{\chi}){\widehat  A}(\Omega^M)
&=&
({\rm dim}\,\chi+
c_1({\mathbb {\widetilde   E}}_{\chi})+
{\rm ch}_2({\mathbb {\widetilde  E}}_{\chi}))
(1-\frac{1}{24}p_1(\Omega^M))
\nonumber \\
& = & {\rm dim}\,\chi+c_1({\mathbb {\widetilde   E}}_{\chi})+
{\rm ch}_2({\mathbb {\widetilde   E}}_{\chi})-
\frac{{\rm dim}\,\chi}{24}p_1(\Omega^M)
\mbox{.}
\end{eqnarray}
The integral over the manifold $M$ takes the form
\begin{equation}
\int_M {\rm ch}({\mathbb {\widetilde E}}_{\chi}){\widehat A}(\Omega^M)
= \int_M {\rm ch}_2({\mathbb {\widetilde E}}_{\chi})-\frac{{\rm dim}\,
\chi}{24}\int_M p_1(\Omega^M)
\mbox{.}
\end{equation}
The following result holds:
\begin{theorem} (M. F. Atiyah, V. K. Patodi and I. M. Singer
\cite{Atiyah1,Atiyah2,Atiyah3}).
\,\,\, The Dirac index is given by
\begin{equation}
{\rm Index}\, {\mathfrak D}_{{\widetilde A}_{\chi}} =\int_{M}{\rm ch}
({\mathbb {\widetilde   E}}_{\chi})
{\widehat  A}(M)-\frac{1}{2}(\eta(0,{\mathfrak D}_{\chi})+
h(0,{\mathfrak D}_{\chi}))
\mbox{,}
\label{Index}
\end{equation}
where $h(0,{\mathfrak D}_{\chi})$ is the dimension of
the space of harmonic
spinors on $X_{\Gamma}$ ($h(0,{\mathfrak D}_{\chi})
={\rm dim}{\rm Ker}\,{\mathfrak D}_{\chi}$ =
multiplicity of the 0-eigenvalue of ${\mathfrak  D}_{\chi}$ acting
on $X_{\Gamma}$); ${\mathfrak D}_{\chi}$ is a Dirac operator on
$X_{\Gamma}$ acting on spinors with coefficients in $\chi$.
\end{theorem}
The Chern-Simons action can be derived from Eq. (\ref{Index}),
\begin{equation}
{\rm Index}\,{\mathfrak D}_{{\widetilde A}_{\chi}} =
CS_{U(n)}({{\widetilde A}_{\chi}})
- \frac{{\rm dim}\,\chi}{24}\int_M p_1(\Omega^M)
- \frac{1}{2}(\eta(0, {\mathfrak D}_{\chi})+
h(0,{\mathfrak D}_{\chi}))
\mbox{.}
\label{I1}
\end{equation}

For type IIA a three form flux associates a phase to a Euclidean
D2-brane world-volume $X_{\Gamma}$ which is given by the eta
invariant of the virtual bundle $\eta_{E_{\chi}} -
\eta_{E_{\chi^{'}}}$ restricted to $X_{\Gamma}$. We can express
this phase directly in terms of the Chern-Simons invariant.
Indeed, for a trivial representation $\chi_0$ one can take a
trivial flat connection $\widetilde{A}_{\chi_0}$; then
$F_{\widetilde{A}_{\chi_0}}=0$ and for this choice we get $ {\rm
Index}\,{\mathfrak D}_{{\widetilde A}_{\chi_0}} = -(1/24)\int_M
p_1(\Omega^M) - (1/2)(\eta(0, {\mathfrak D})+ h(0,{\mathfrak D}))
$. Using this formula in (\ref{I1}) one can obtain
\begin{equation}
{\rm Index}\,{\mathfrak D}_{{\widetilde A}_{\chi}} - {\rm
dim}{\chi}\,{\rm Index}\,{\mathfrak D}_{{\widetilde A}_{\chi_0}} =
CS_{U(n)}({{\widetilde A}_{\chi}}) - \frac{1}{2}(\eta(0,
{\mathfrak D}_{\chi}) - {\rm dim}{\chi}\eta(0, {\mathfrak D}))
\,\,\, {\rm modulo}({\mathbb Z}/2) \mbox{.}
\end{equation}
Finally we get the following result for the invariant of an
irreducible flat connection on the real hyperbolic three-manifolds
\begin{eqnarray}
\frac{1}{2}\left({\rm dim}{\chi}\,\eta(0,{\mathfrak D})-
\eta(0,{\mathfrak D}_{\chi})\right) && = \frac{1}{2\pi {\sqrt{-1}}}{\rm log}
\left[\frac{Z(0,{\mathfrak D})^{{\rm dim}\chi}} {Z(0,{\mathfrak
D}_{\chi})}\right] \nonumber \\
&&=CS_{U(n)}({\widetilde
A}_{\chi})\,\,\,  {\rm modulo}({\mathbb Z}/2)
 \mbox{,} \label{CSF}
\end{eqnarray}
since the index of a Dirac operator,
acting on some spin four-manifold, is an integer.

\section{Dirac operators and K-theory class}

In the previous section we have calculated the Chern character,
the Chern classes and the Chern invariants of (3d) hyperbolic
geometry, using the fact that the index of a Dirac operator,
acting on some spin four-manifold, is an integer. Here we would
like to give explicit formulas for the index of a Dirac operator
(and twisted Dirac operator) on an even dimensional manifold.
First we need to explain the intimate connection between the index
theory of Fredholm operators and topological K-theory. Let
${\mathcal O}$ be a {\it Fredholm operator}, acting on a separable
Hilbert space ${\mathcal H}$. It is a bounded linear operator
whose kernel (and cokernel) are finite dimensional subspaces of
$\mathcal H$. Such operators have a well-defined {\it Index},
which is invariant under perturbations by any compact operator
${\mathcal A}$:
\begin{eqnarray}
&& {\rm Index}\,({\mathcal O} + {\mathcal A}) = {\rm
Index}\,({\mathcal O}); \nonumber
\\
&& {\rm Index}\,({\mathcal O}) =  {\rm Index}\,{\mathcal A}\,\,\,
{\rm if}\,\,\, {\mathcal A}\,\,\, {\rm is}\,\,\, {\rm
sufficiently} \,\,\, {\rm close}\,\,\, {\rm in}\,\,\, {\rm
the}\,\,\, {\rm operator} \,\,\, {\rm norm}\,\,\, {\rm to}\,\,\,
{\mathcal O} \nonumber
\mbox{.}
\end{eqnarray}
These formulas are important since one can describe the group
$K(X)$ (and groups of cohomology) in terms of Fredholm operators.
Indeed, let ${\mathfrak X}$ be the space of Fredholm operators on
${\cal H}$ with the operator norm topology. Then the index of a
Fredholm operator defines a continuous map ${\rm Index}\,:\,
{\mathfrak X}\rightarrow {\mathbb Z}$ which induces a bijection
\cite{Olsen}: $ \pi_0({\mathfrak X})\rightarrow {\mathbb Z} $
between the set of connected components of ${\mathfrak X}$ and the
integers. Let $X$ be a {\it compact topological space}. Suppose
$[X,{\mathfrak X}]$ be a set of homotopy classes of maps from $X$
to ${\mathfrak X}$. The product of two Fredholm operators is again
a Fredholm operator, which implies that $\left[X,{\mathfrak
X}\right]$ is a monoid. There is an isomorphism \cite{Olsen}: $
\left[X,{\mathfrak X}\right]\stackrel{\approx}{\rightarrow} K(X).
$ Let $\{{\mathcal O}\}_{x\in X}$ be a continuous family of
Fredholm operators. The family of vector spaces ${\rm
Ker}\,{\mathcal O}_x$ (${\rm coKer}\,{\mathcal O}_x)$ forms a
vector bundle ${\rm Ker}\,{\mathcal O}$ ( ${\rm coKer}\,{\mathcal
O}$) over $X$, and we can define the index of a family of
operators ${\mathcal O}_x$ as the class $ {\rm Index}\,{\mathcal
O}\equiv [({\rm Ker}\,{\mathcal O}\,,\, {\rm coKer}\,{\mathcal
O})]\in K(X)\ . $ Thus, the composition of operators in ${\mathcal
O}$ corresponds to the addition in $K(X)$, while adjoint operation
corresponds to inversion. In the case where $X$ is a point, $K(X =
pt)= {\mathbb Z}$, the isomorphism $ \left[pt,{\mathfrak
X}\right]\stackrel{\approx}{\rightarrow} {\mathbb Z} $ is just the
index map $ \pi_0({\mathfrak X})\rightarrow {\mathbb Z}, $ the
virtual dimension of the K-theory class coincides with the index
of a Fredholm operator, i.e. : $ {\rm ch}_0({\rm Index}\,{\mathcal
O}) = {\rm Index}\,{\mathcal O} \ . $

We are interested in applying these ideas to a special class of
operators, the Dirac operators associated to vector bundles over a
spin manifold $X$ (a standard material on Dirac bundles the reader
can find for example in the book \cite{Laws89}). Dirac operators
are examples of pseudo-differential elliptic operators, which are
Fredholm operators when viewed as operators on a Hilbert space.

\subsection{The index of Dirac operator}

Taking into account our interest in hyperbolic geometry, we
compute in this section the index of Dirac operator acting on real
closed hyperbolic manifold (which is in connection with
Chern-Simons and eta topological invariants, as we have seen in
previous section). We actually consider the special case
$X=G/{\mathcal K}$, $G=SO_1(2n,1)$, \, ${\mathcal K} =SO(2n)$. The
complexified Lie algebra ${\mathfrak g}={\mathfrak g}^{\mathbb
C}_0 ={\mathfrak so}(2n+1,{\mathbb C})$ of $G$ is of the Cartan
type $B_n$. Let ${\mathfrak a}_0, {\mathfrak n}_0$ denote the Lie
algebras of $A, N$ in an Iwasawa decomposition $G={\mathcal K}AN$.
Since the rank of $G$ is one, $\dim {\mathfrak a}_0=1$ by
definition, say ${\mathfrak a}_0={\mathbb R}H_0$ for a suitable
basis vector $H_0$: $H_0 = {\rm antidiag}(1, 1, ..., 1)$. By this
choice we have the normalization $\beta(H_0)=1$, where $\beta:
{\mathfrak a}_0\rightarrow{\mathbb R}$ is the positive root which
defines ${\mathfrak n}_0$.
The standard systems of positive roots $\triangle^{+},\triangle^{+}_\ell$
for ${\mathfrak g}$ and ${\mathfrak k}={\mathfrak k}^ {\mathbb C}_0-$
the complexified Lie algebra of ${\mathcal K}$, with
respect to a Cartan subgroup $H$ of $G$,\, $H\subset {\mathcal K}$,
are given by
$
\triangle^{+}=\{\varepsilon_i|1\leq i\leq n\}\bigcup
\triangle^{+}_\ell\,,
$
where
$
\triangle^{+}_\ell=\{\varepsilon_i\pm \varepsilon_j|1\leq i<j\leq n\},
$
and
$
\triangle^{+}_n \stackrel{def}{=}\{\varepsilon_i|1\leq i\leq n\}
$
is the set of positive non-compact roots. Here
$
(\varepsilon_i,\, \varepsilon_j) = \delta_{ij}[(H_0, H_0)]^{-1} =
\delta_{ij} [2(2n-1)]^{-1}\,,
(\varepsilon_i \pm \varepsilon_j,\,\,\varepsilon_i \pm \varepsilon_j)
= [2n+1]^{-1}\,,\,\,\, i<j\,;
$
i.e.
$
(\alpha, \alpha) =  [2n-1]^{-1}\,,
\forall \alpha\in \triangle^{+}_n\,.
$

Let ${\mathfrak h}_0$
be the Lie algebra of $H$ and let
${\mathfrak h}^{*}_{\mathbb R}={\rm Hom}(\sqrt{-1}{\mathfrak h}_0,
{\mathbb R})$ be the dual space of the real vector space
$\sqrt{-1}{\mathfrak h}_0$. Thus, the
$\{\varepsilon_i\}_{i=1}$ are an ${\mathbb R}$-basis
of ${\mathfrak h}^{*}_{\mathbb R}$. Of interest are the
{\em integral} elements
$f$ of ${\mathfrak h}^{*}_{\mathbb R}$:
\begin{eqnarray}
f\stackrel{def}{=}\{\mu\in {\mathfrak h}^{*}_{\mathbb R}|
\frac{2(\mu,\alpha)}
{(\alpha,\alpha)}\in
{\mathbb Z},\,\,\, \forall \alpha\in \triangle^{+}\}
\mbox{.}
\end{eqnarray}
Then we have
$
2(\mu,\varepsilon_i)[(\varepsilon_i,\varepsilon_i)]^{-1}
= 2\mu_i\,,
$
for $1\leq i\leq n$,
$
2(\mu,\varepsilon_i\pm \varepsilon_j)
[(\varepsilon_i\pm \varepsilon_j, \varepsilon_i\pm \varepsilon_j)]^{-1}
= \mu_i\pm \mu_j\,,
$
for $1\leq i<j\leq n$, where $\mu=\sum_{j=1}^n\mu_j\varepsilon_j$
for $\mu\in {\mathfrak h}^{*}_{\mathbb R},\, \mu_j\in {\mathbb R}$.
Let
$
\delta_k= (1/2)\sum_{\alpha\in\triangle^{+}_k}\alpha, \,\,
\delta_n= (1/2)\sum_{\alpha\in\triangle^{+}_n}\alpha, \,\,
\delta=\delta_k+\delta_n= (1/2)\sum_{\alpha\in\triangle^{+}}\alpha\,,
$
then
$
\delta_k=\sum_{i=1}^n(n-i)\varepsilon_i,\,\,
\delta_n= (1/2)\sum_{i=1}^n\varepsilon_i,\,\,
\delta=\sum_{i=1}^n(n-i- 1/2)\varepsilon_i
$
are all integral.
The elements $\mu$ of $f$ correspond to characters $e^{\mu}$ of $H$.
Following \cite{Barbasch} we fix, once and for all, a $\mu\in f$,
which is $\triangle^{+}_k$-dominant: $(\mu,\alpha)\geq 0$ for
$\alpha\in\triangle^{+}_k$. For us, in concrete terms, this means the
following. Let $\mu=(\mu_1,...,\mu_n)$ be a sequence of real numbers
such that
\begin{eqnarray}
&& (1)\,\, 2\mu_i\in {\mathbb Z}\,\,\, {\rm for}\,\,\, 1\leq i\leq n
\,\,\,\,{\rm and}\,\,\,\,
\mu_i\pm \mu_j \in {\mathbb Z}\,\,\,\, {\rm for}\,\,\, 1\leq i<j<n
\,\,\, ({\rm i.e.}\,\, \mu \,\, {\rm is}\,\, {\rm integral}),
\nonumber\\
&& (2)\,|\mu_n|\leq\mu_{n-1}\leq\mu_{n-2}\leq...\leq\mu_2\leq\mu_1
\,\,\,\,{\rm and}\,\,\,\, {\rm either}\,\,\,\, {\rm every}
\,\,\,\, \mu_i\in {\mathbb Z}\,\,\,{\rm or},\,\, {\rm else},
\nonumber \\
&& \,\,\,\,\,\,\,\,\,\,
{\rm every}\,\,\,
\mu_i\,\,\, {\rm is}\,\,\, {\rm half}\,\,\, {\rm integer}.
\nonumber
\end{eqnarray}
Note that, in fact, $(1)\rightarrow (2)$ so that we may drop the
condition $(1)$ (only for $G=SO_1(2n,1)$ since in general $(2)$
does not imply $(1)$). Then $\mu+\delta_n$ does define a character
of $H$, as has been required in \cite{Barbasch} and, by
construction, there is a twisted Dirac operator ${\mathfrak
D}_{\mu,\Gamma}$ on a vector bundle over $\Gamma\backslash
G/{\mathcal K}$ for $\Gamma$ a discrete subgroup of $G$. Thus,
$\Gamma\backslash G$ does not need to be compact, but one requires
of course that ${\rm Vol}(\Gamma\backslash G)<\infty$.

{\it Twisted Dirac operator}.
In summary, we assume the following:
$\mu=(\mu_1,...,\mu_n)$ is a sequence of real numbers, and
\begin{eqnarray}
&& (3)\,\,\, 0\leq\mu_n\leq\mu_{n-1}\leq\mu_{n-2}
\leq\cdot\cdot\cdot\leq\mu_2\leq\mu_1,\,\,\, {\rm where}\,\,\,
{\rm either}\,\,\, {\rm every}\,\,\, \mu_j\in {\mathbb Z}
\nonumber \\
&& \,\,\,\,\,\,\,\,\,\, {\rm or}\,\,\, {\rm every}\,\,\,
\mu_j\,\,\, {\rm has}\,\,\, {\rm the}\,\,\, {\rm form}\,\,\,
\mu_j=n_j+ 1/2\,\,\, {\rm for}\,\,\, {\rm some}\,\,\,
n_j\in {\mathbb Z}.
\nonumber
\end{eqnarray}
Then we have a twisted Dirac operator ${\mathfrak D}_{\mu,\Gamma}$
over $\Gamma\backslash G/{\mathcal K}$.
\begin{theorem} (\cite{Barbasch}, for the case $G=SO_1(2n,1),
n\geq 2$).\,\,\,
For a suitable normalization of the Haar measure on $G$,
and for $\mu$ satisfying the condition (3), one has
\begin{equation}
{\rm Index}\,{\mathfrak D}_{\mu,\Gamma}={\rm Vol}(\Gamma\backslash G)
\frac{\prod_{\alpha\in \triangle^{+}}(\mu+\delta_k,\alpha)}
{\prod_{\alpha\in \triangle^{+}_k}(\delta_k,\alpha)}
\mbox{.}
\label{Ind1}
\end{equation}
\end{theorem}
To make this explicit, we must express $P_k\stackrel{def}{=}
\prod_{\alpha\in\triangle^{+}_k}(\delta_k,\alpha)$ and
$P\stackrel{def}{=}\prod_{\alpha\in\triangle^{+}}(\mu+\delta,\alpha)$
directly
in terms of the real numbers $\mu_1,...,\mu_n$.
First we compute $P_k\stackrel{def}{=}\prod_{\alpha\in\triangle^{+}_k}
(\delta_k,\alpha)$.
Note that the number of pairs $(i,j)$ with $1\leq i<j\leq n$ is
$n(n-1)/2$;
$
(\delta_k,\delta_j) =
\sum_{i=1}^n(n-i)(\varepsilon_i,\varepsilon_j)
= [n-j][2(2n-1)]^{-1}
\Longrightarrow
(\delta_k,\varepsilon_i\pm \varepsilon_j) =
[n-i\pm (n-j)][2(2n-1)]^{-1}\,,
$
or
\begin{eqnarray}
P_k &=& \prod_{1\leq i<j\leq n}(\delta_k,\varepsilon_i+\varepsilon_j)
\prod_{1\leq i<j\leq n}(\delta_k,\varepsilon_i-\varepsilon_j)
= \prod_{1\leq i<j\leq n}\frac{2n-i-j}{2(2n-1)}
\prod_{1\leq i<j\leq n}\frac{-i+j}{2(2n-1)}
\nonumber \\
&=&\left[\frac{1}{2(2n-1)}\right]^{n(n-1)}
\prod_{1\leq i<j\leq n}(2n-i-j)(-i+j)
\mbox{.}
\end{eqnarray}
We also look at $P\stackrel{def}{=}\prod_{\alpha\in\triangle^{+}}
(\mu+\delta_k,\alpha)$;
$
\mu+\delta_k =  \sum_{j=1}^n(\mu_j+n-j)\varepsilon_j\,\,\,\,\,\,\,
\Longrightarrow
(\mu+\delta_k,\varepsilon_i) =  \sum_{j=1}^n(\mu_j+n-j)(\varepsilon_j,
\varepsilon_i) = (\mu_i+n-i)[2(2n-1)]^{-1}\,.
$
Then, for $i<j$,
$
(\mu+\delta_k,\varepsilon_i+\varepsilon_j) =
(\mu_i+\mu_j+2n-i-j)[2(2n-1)]^{-1}\,,
$
$
(\mu+\delta_k,\varepsilon_i-\varepsilon_j) =
(\mu_i-\mu_j+i-j)[2(2n-1)]^{-1}\,,
$
and
\begin{eqnarray}
&& \prod_{\alpha\in\triangle^{+}_k}(\mu+\delta_k,\alpha) =
\prod_{1\leq i\leq j\leq n}\frac{\mu_i+\mu_j+2n-i-j}{2(2n-1)}
\prod_{1\leq i\leq j\leq n}\frac{\mu_i-\mu_j-i+j}{2(2n-1)}
\nonumber\\
&& =\left[\frac{1}{2(2n-1)}\right]^{n(n-1)}
\prod_{1\leq i\leq j\leq n}(\mu_i+\mu_j+2n-i-j)(\mu_i-\mu_j-i+j)
\mbox{.}
\end{eqnarray}
Thus we have
\begin{equation}
\prod_{\alpha\in\triangle^{+}_n}(\mu+\delta_k,\alpha)=
\prod_{i=1}^n\frac{(\mu_i+n-i)}{2(2n-1)}=
\left[\frac{1}{2(2n-1)}\right]^n\prod_{i=1}^n(\mu_i+n-i)
\mbox{,}
\label{1}
\end{equation}
and therefore,
\begin{equation}
P = \left[\frac{1}{2(2n-1)}\right]^{n^2}\prod_{i=1}^n(\mu_i+n-i)
\prod_{1\leq i<j\leq n}(\mu_i+\mu_j+2n-i-j)(\mu_i-\mu_j-i+j)
\mbox{.}
\label{2}
\end{equation}
We arrive at the final result for $\mu=(\mu_1,...,\mu_n)\in
{\mathbb R}^n$, subject to the condition (3) (for a suitable Haar
measure on $G$). The final result follows from formulas
(\ref{Ind1}), (\ref{1}) and (\ref{2}). The $L^2$-index of the
twisted Dirac operator ${\mathfrak D}_{\mu,\Gamma}$ is equal to
\begin{eqnarray}
{\rm Index}\,{\mathfrak D}_{\mu,\Gamma}& = &
{\rm Vol}(\Gamma\backslash G)
\frac{P}{P_k}=
\frac{{\rm Vol}(\Gamma\backslash G)}{[2(2n-1)^n]}
\prod_{i=1}^n (\mu_i+n-i)
\nonumber \\
& \times &
\frac{\prod_{1\leq i<j\leq n}(\mu_i+\mu_j+2n-i-j)
(\mu_i-\mu_j-i+j)}{\prod_{1\leq i<j\leq n}(2n-i-j)(-i+j)}
\mbox{.}
\label{Ind2}
\end{eqnarray}

\section{Brane charges, KK-groups and K-amenable groups}

In many instances of field and string theory, torsion can make
its appearance in the integral cohomology groups $H^\#(X,{\mathbb Z})$.
But local fields are represented by differential forms,
consequently by themselves they cannot carry torsion.
In previous sections, in particular, we saw that the
Chern isomorphism determines the free part of the K-groups from
ordinary cohomology -- a well-known result, which allows us
the avoid more sophisticated mathematical concepts. We also saw that
in some fortunate circumstances we can express torsion charges by means
of the $\eta$ invariant. However in general, for suitable spaces
$X$ (for example, topological manifolds or
finite CW- complexes) $K(X)$ is a finitely generated abelian
group, it has the form ${\mathbb Z}^{\ell}\oplus {\rm Tor}$, and
in order to fully determine RR--fields and charges
we have to resort to full-fledged K-theory.

In this section we would like to broach this subject as far as
K-theory on hyperbolic spaces is concerned. This requires the
introduction of new mathematical tools: K-homology, KK-theory and
K-amenabilty, which will allow us to state some results on twisted
K-theory on hyperbolic spaces.

\subsection{KK-groups}

Given a manifold $X$ let $C(X)$ be the commutative $C^*-$algebra
\footnote{Recall that a $C^*-$algebra is a Banach algebra with an
involution satisfying the relation $||a^*||=||a||^2$.} of all
continuous complex-valued functions which vanish at infinity on
$X$. The $C^*-$algebra, which categorically encodes the
topological properties of manifold $X$, plays a dual role to $X$ in the
K-theory of $X$ by the Serre-Swan theorem \cite{WO}:
$K^\ell(X)\cong K_{\ell}(C(X)), \ell = 0, 1$.\, Here $K^\ell(X)$
is the reduced topological K-theory of $X$ (see section Appendix
for details). Let $X$ have a $Spin^{\mathbb C}-$structure, then
there is a Poincar\'{e} duality isomorphism \cite{Higson}:
$K^{{\rm dim}\,X-\ell}(X)\cong K_{\ell}^{\mathbb C}(X),\, \ell =0,
1$, where $K_{\ell}^{\mathbb C}$ denotes the dual compactly
supported K-homology of $X$.

For a finite dimensional manifold $X$ the exists another
$C^*-$algebra, which is non-commutative and can be constructed
with the help of the Riemannian metric $g$. In fact, we can form
the complex Clifford algebra ${\rm Cliff}(T_xX,g_x)$, where for
each $x\in X$ the tangent space $T_xX$ of $X$ is a
finite-dimensional Euclidean space with inner product $g_x$. This
algebra has a canonical structure as a finite-dimensional
${\mathbb Z}_2-$graded $C^*-$algebra. Let us consider the family
of $C^*-$algebras $\{{\rm Cliff}(T_xX, g_x\}_{x\in X}$; it forms a
${\mathbb Z}_2-$graded $C^*-$algebra vector bundle ${\rm
Cliff}(TX)\rightarrow X$, called the Clifford algebra bundle of
$X$ \cite{Atiyah64}. Let us define ${\mathcal C}(X) = C(X, {\rm
Cliff}(TX))$ to be the $C^*-$algebra of continuous sections of the
Clifford algebra bundle of $X$ vanishing at infinity. If the
manifold $X$ is even-dimensional and has a $Spin^{\mathbb
C}-$structure then this $C^*-$algebra is Morita equivalent to
$C(X)$. However, in general, ${\mathcal C}(X)$ is Morita
equivalent to $C(TX)$. Because of the Morita equivalence of
K-theory, it follows that $K_{\ell}({\mathcal C}(X))\cong
K_{\ell}(C(X))\cong K^{\ell}(X), \ell =0, 1$. But for
odd-dimensional and spin manifold $X$ this relation is more
complicated.

{\it KK-pairing}. Let us recall that the definition of K-homology
involves classifying extensions of the algebra of continuous
functions $C(X)$ on the manifold $X$ by the algebra of compact
operators up to unitary equivalence \cite{Brown}. The set of
homotopy classes of operators defines the K-homology group
$K_0(X)$, and the duality with K-theory is provided by the natural
bilinear pairing $ ([E]\,,\,[{\mathfrak D}])\mapsto {\rm Index}\,
{\mathfrak D}_{E}\in {\mathbb Z}\,, $ where $[E]\in K(X)$ and
${\mathfrak D}_E$ denotes the action of the Fredholm operator
${\mathfrak D}$ on the Hilbert space ${\mathcal H}= L^2(U(X,E))$
of square-integrable sections of the vector bundle $E\rightarrow
X$ as ${\mathfrak D}: U(X,E)\rightarrow U(X,E)$. Hence, it assumes
that the KK-pairing may be the most natural framework for our
purpose. The KK-theory is a bivariant version of topological
K-theory. It provides a useful framework for the study of
the index theory.

The main object here is the group $KK(A,B)$, which depends on a
pair of graded algebras
$A$ and $B$. Let $A$ and $B$ be $C^*-$algebras.
A pair $({\mathcal E},
\pi)$, where ${\mathcal E}$ is a ${\mathbb Z}/2{\mathbb Z}$ graded
Hilbert $B-$module acted upon by $A$ through a $*-$homomorphism
$\pi: A\rightarrow {\mathcal L}({\mathcal E}) = {\rm End}^*({\mathcal E})$,
$\forall a\in A$ the operator $\pi (a)$ being of degree $0, \pi (A)
\subset {\mathcal L}({\mathcal E})^{(0)}$, will be called an $(A,
B)-$bimodule. Let $E(A,B)$ be a triple $({\mathcal E}, \pi , F)$,
where $({\mathcal E}, \pi)$ is a $A,B-$module, $F\in {\mathcal
L}({\mathcal E})$ is a homogeneous operator of degree 1, and
$\forall a \in A$:
\begin{eqnarray}
&& ({\rm I})\,\,\,\,\,\, \pi (a)(F^2-1)\in {C}({\mathcal E}),
\nonumber \\
&& ({\rm II})\,\,\, [\pi (a), F] \in {C}({\mathcal E})\,\,
\nonumber
\end{eqnarray}
where ${C}({\mathcal E})$ is the algebra of compact operators.
A triple $({\mathcal E}, \pi , F)$ will be called degenerate if
$\forall a \in A$: $\pi (a)(F^2-1) = 0$, $[\pi (a), F] = 0$. Let
$D(A, B)$ be a set of generated triples. An element $E(A, B[0,1])$,
where $B[0,1]$ is an algebra of continuous functions in $B$ on the
interval $[0,1]$, will be called a homotopy in $E(A,B)$. Let us
assign a direct sum in $E(A,B)$:
$$
({\mathcal E}, \pi ,
F)\oplus({\mathcal E}^{'}, \pi^{'} , F^{'}) = ({\mathcal
E}\oplus{\mathcal E}^{'},\pi\oplus\pi^{'},F\oplus F^{'})\,.
$$
The homotopy classes of $E(A,B)$ together with this sum define
the Abelian group $KK(A,B)$.
A $*-$homomorphism, $f: A_1\rightarrow A_2$, transfers
$(A_2, B)-$modules into $(A_1, B)-$ modules, and \cite{Kasparov1}
$$
f^*: E(A_2,B)\rightarrow E(A_1,B)\,,\,\,
({\mathcal E},\pi,F)\mapsto ({\mathcal E},\pi^{\circ}f,F)\,,
$$
On the other hand a $(*)-$homomorphism $g:
B_1\rightarrow B_2$ induces a homomorphism
$$
g_*: \, E(A,B_1)\rightarrow E(A,B_2)\,,\,\,\,\,
({\mathcal E},\pi,f) \mapsto
({\mathcal E}\otimes_gB_2,\pi\otimes 1,F\otimes 1)\,,
$$
where
$$
\pi\otimes 1: A\rightarrow {\mathcal L}({\mathcal E}\otimes_gB_2)\,,
\,\,\,\,\,\,\,
(\pi\otimes 1)(a)(e\otimes b)=\pi(a)e\otimes b\,.
$$
\begin{theorem}
\,\,\, The groups
$KK(A,B)$ define a homotopy invariant bifunctor from the category of
separable $C^*-$algebras into the category of Abelian groups.
Abelian groups $KK(A,B)$ depend covariantly on the algebras $A$
and $B$, in addition $KK({\mathbb C}, B)= K_0(B)$.
\end{theorem}
The facts of interest to us are the following relations:
\begin{equation}
K\!K^{*}({A:=\mathbb C}, B) = K_{*}(B), \,\,\,\,\,\,\,
K\!K^{*}(A, B:={\mathbb C}) = K^{*}(A)
\mbox{.}
\end{equation}
\begin{definition}\,\,\,
Let  $1_A \in KK(A,A)$ ($KK(A,A)$ is a ring with unit) denotes the
triple class $(A, \iota_A,0)$ where $A^{(1)}=A,\, A^{(0)}=0$, and
$\iota_A: A\rightarrow {\mathcal K}(A)\subset {\mathcal L}(A)$,
$\iota_A(a)b = ab,\, a,b\in A$.
\end{definition}

Let us define also the map
\begin{eqnarray}
&&
\tau_D : \,\,
KK(A,B)\otimes KK(A\otimes D, B\otimes D)\,,
\nonumber \\
&&
\tau_D\,({\rm class}\,\,({\mathcal E}, \pi, F)) =
{\rm class}\,\, ({\mathcal E}\otimes D, \pi \otimes 1_D, F \otimes 1)\,.
\nonumber
\end{eqnarray}

The most important object of our concern is
Kasparov's pairing
\begin{theorem}\,\,\,Kasparov's pairing, defined by
\begin{equation}
KK(A,D)\times KK(D,B)\longrightarrow KK(A,B)
\end{equation}
and denoted $(x,y)\mapsto x\otimes_Dy$,
satisfies the following properties:
\begin{itemize}
\item{} It depends covariantly on
the algebra $B$ and contravariantly on the algebra $A$.
\item{} If $f: D\rightarrow E$ is a $*-$homomorphism, then
$f_*(x)\otimes_Ey =x\otimes_Df^*(y)$,\,
$x\in KK(A,D),\, y\in KK(E, B)$.
\item{} Associative property:
$(z\otimes_Dy)\otimes_Ez=x\otimes_D(y\otimes)_Ez$,\,
$\forall x\in KK(A,D),\, y\in KK(D,E),\, z\in KK(E,B)$.
\item{} $x\otimes_B1_B=1_A\otimes x =x,\, \forall x\in KK(A,B)$.
\item{} $\tau_E(x\otimes_{B} y)=\tau_E(x)\otimes_{B\otimes E}\tau_E(y)\,,
\,\, \forall x\in KK(A, B),\, \forall y \in KK(B,D)$.
\end{itemize}
\end{theorem}

Suppose that for two algebras, $A$ and $B$, there are elements
$\alpha\in K\!K(A\otimes B,{\mathbb C})$, $\beta\in K\!K({\mathbb
C},A\otimes B)$, with the property that $\beta\otimes_{A}\alpha =
1_{B} \in\ K\!K(B,B)$,  $\beta\otimes_{B}\alpha = 1_{A} \in\
K\!K(A,A).$ Then we say we have KK-duality isomorphisms between
the K-theory (K-homology) of the algebra $A$ and the K-homology
(K-theory) of the algebra $B$
$$
K_{*}(A) \cong K^{*}(B) ,\,\,
K^{*}(A) \cong K_{*}(B)\,.
$$
In fact the algebras $A$ and $B$ are Poincar\'e dual
\cite{Connes94}, but generally speaking these algebras are not
KK-equivalent.

{\it K-amenable groups}. We now review the concept of K-amenable
groups \cite{Carey}. Let $G$ be a connected Lie group and
${\mathcal K}$ a maximal compact subgroup. We also assume that
${\rm dim}\, (G/{\mathcal K})$ is even and $G/{\mathcal K}$ admits
a $G-$invariant ${\rm Spin}^{\mathbb C}$ structure. The
$G-$invariant Dirac operator ${\mathfrak D :=
\gamma^{\mu}\partial_{\mu}}$ on $G/{\mathcal K}$ is a first order
self-adjoint, elliptic differential operator acting on $L^2-$
sections of the ${\mathbb Z}_2-$graded homogeneous bundle of
spinors ${\mathcal S}$. Let us consider a $0^{\rm th}$ order
pseudo-differential operator ${\mathcal O} = {\mathfrak
D(1+{\mathfrak D}^2)^{-\frac 12}}$ acting on $H=L^2(G/{\mathcal
K}, {\mathcal S})$. $C(G/{\mathcal K})$ acts on $H$ by
multiplication of operators. $G$ acts on $C(G/{\mathcal K})$ and
on $H$ by left translation, and ${\mathcal O}$ is $G-$invariant.
Then, the set $({\mathcal O}, H, X)$ defines a canonical Dirac
element $\alpha_G = KK_G(C(G/{\mathcal K}), {\mathbb C})$.
\begin{theorem} (G. Kasparov \cite{Kasparov95})\,\,\,
There is a canonical Mishchenko element
$$
\alpha_G \in KK_G(C(G/{\mathcal K}), {\mathbb C})
$$
such that the following intersection products occur:
\begin{itemize}
\item{} $\alpha_G\otimes_{\mathbb C}\beta_G = 1_{C(G/{\mathcal
K})} \in KK_G(C(G/{\mathcal K}),C(G/{\mathcal K}))$\,, \item{}
$\beta_G\otimes_{C(G/{\mathcal K})}\alpha_G=\gamma_G=
KK_G({\mathbb C}, {\mathbb C})$ where $\gamma_G$ is an element in
$KK_G({\mathbb C}, {\mathbb C})$.
\end{itemize}
\end{theorem}
For a semisimple Lie group $G$ or for $G={\mathbb R}^n$, a
construction of the Mishchenko element $\beta_G$ can be found in
\cite{Carey}. We now come to the basic definition:
\begin{definition}\,\,\, A Lie group $G$ is said to be
K-amenable if $\gamma_G$=1.
\end{definition}
We recall that a group $G$ is amenable if there exists a left
invariant positive linear functional on the space $B(G)$ of all
continuous bounded functions on $G$ with the norm $\parallel f
\parallel= {\rm sup} |f{x}|$. All solvable groups are amenable,
while any non-compact semisimple Lie group is non--amenable.
However the non-amenable groups $SO_1(n,1)$ and $SU(n,1)$ are
K-amenable \cite{Kasparov,Julg}.

\subsection{Twisted cohomology and K-theory}.

In the presence of a nontrivial background B-field the
classification of charges and RR--fields needs a radical
modification. It was argued by Witten, \cite{Witten}, that, when
$H$, the curvature of $B$, is pure torsion (flat $B$), the D-brane
charges take value in a twisted version of K-theory. Subsequently
in  \cite{Bouwknegt}, this construction was extended to the case
$[H]\neq 0$. In this case it was shown that brane charges  are
classified by certain infinite dimensional algebra bundles of
compact operators, introduced by Dixmier and Douady,
\cite{Dixmier}. More in detail, it was proposed that in the
presence of a non-flat B-field D-brane charges in type IIB string
be measured by the twisted K-theory that has been described by
Rosenberg, \cite{Rosenberg}, and the twisted bundles on the
D-brane world-volumes be elements in this twisted K-theory.

First we would like to note the result given in \cite{Elliott,Connes}:
\begin{equation}
K_\#(C^*({\mathbb Z},\sigma))\cong
K_\#(C^*({\mathbb Z}^n)) \cong K_\#({\mathbb T}^n)\,,
\end{equation}
which holds for any group two-cocycle $\sigma$ on ${\mathbb Z}^n$.
This calculation leads to the twisted group of $C^*-$algebras
$C^*({\mathbb Z},\sigma)$ (noncommutative tori). Such
generalization has been given for K-groups of the twisted group of
$C^*-$algebras of uniform lattice in solvable groups
\cite{Packer1}. Let $\Gamma$ be a lattice in a K-amenable Lie
group $G$, then \cite{Carey}:
\begin{equation}
K_\#(C^*({\Gamma},\sigma))\cong K^{\#+{\rm dim}(C/{\mathcal K})}
(\Gamma\backslash G/{\mathcal K}, \delta(B_{\sigma}))
\mbox{,}
\end{equation}
where $K^{\#+{\rm dim} G}(\Gamma\backslash G, \delta(B_{\sigma}))$
denotes the twisted K-theory (see \cite{Rosenberg}) of a
continuous trace $C^*-$algebra $B_{\sigma}$ with spectrum
$\Gamma\backslash G$, $\sigma$ is any multiplier on $\Gamma$,
while $\delta (B_{\sigma}) \in H^3(\Gamma\backslash G, {\mathbb
Z})$ denotes the Dixmier-Douady invariant of $B_{\sigma}$
\cite{Dixmier}. Let $\Gamma$ be a lattice in a K-amenable Lie
group $G$. Then the following formulas hold \cite{Carey}:
\begin{eqnarray}
K_\#(C^*({\Gamma},\sigma))
& \cong &
K^{\#+{\rm dim} G}(C^*_r(\Gamma, \sigma))\,,
\\
K_\#(C^*({\Gamma},\sigma))
& \cong &
K^{\#+{\rm dim} G/{\mathcal K}}(\Gamma\backslash G/{\mathcal K},
\delta(B_{\sigma}))
\mbox{.}
\end{eqnarray}
These formulas  have been obtained with the help of K-amenability
results \cite{Kasparov} and the stabilization theorem
\cite{Packer1}. When $\Gamma = \Gamma_g$ is a fundamental group of
a Riemannian surface $X_{\Gamma} = \Sigma_g$ of genus $g>0$ the
Dixmier-Douadly class $\delta(B_{\sigma})$ is trivial and we get
\begin{equation}
K_0(C^*({\Gamma}_g,\sigma)) \cong
K^{0}(\Sigma_g) \cong {\mathbb Z}^2\,,\,\,\,\,\,
K_1(C^*({\Gamma}_g,\sigma)) \cong
K^{1}(\Sigma_g) \cong {\mathbb Z}^{2g}
\mbox{,}
\end{equation}
which hold for any multiplier $\sigma$ on $\Gamma_{\sigma}$.

\section{Lower K-groups}

In this final section we collect a few more basic results on the lower
algebraic K-groups of hyperbolic spaces. Let $X_{\Gamma}$ be a
real compact oriented three dimensional hyperbolic manifold. Its
fundamental group $\Gamma$ comes with maps to $PSL(2, {\mathbb
C})\equiv SL(2, {\mathbb C})/\{\pm Id\}$; therefore in general one
gets a class in $H_3(GL({\mathbb C}))$. The following result holds
\cite{Suslin1,Suslin3}: For a field ${\mathcal F}$,
\begin{equation}
K_\ell({\mathcal F})\cong H_\ell(GL(\ell, {\mathcal
F}))/H_\ell(GL(\ell-1, {\mathcal F}) \mbox{.}
\end{equation}
The group $K_3({\mathcal F})$ is built out of $K_3({\mathcal F})$
and the Bloch group ${\mathfrak B}({\mathcal F})$. \footnote{
There is an exact sequence due to Bloch and Wigner:
$$
0\rightarrow {\mathbb Q}/{\mathbb Z}\rightarrow H_3(PSL(2,
{\mathbb C}); {\mathbb Z})\rightarrow {\mathfrak B}({\mathbb
C})\rightarrow 0
$$
The Bloch group ${\mathfrak B}({\mathbb C})$ is known to be
uniquely divisible, so it has canonically the structure of a
${\mathbb Q}-$vector space \cite{Suslin87}. The ${\mathbb
Q}/{\mathbb Z}$ in the Bloch-Wigner exact sequence is precisely
the torsion of $H_3(PSL(2, {\mathbb C}); {\mathbb Z})$.}
 Since we are
looking for $H_3(GL(2, \bullet))$, the homology invariant of a
hyperbolic three-manifold should live in the Bloch group
${\mathfrak B}(\bullet)$. The following result confirmed that
statement \cite{Neumann1}: A real oriented finite-volume
hyperbolic three-manifold $X = X_{\Gamma}$ has the Bloch invariant
$\beta (X) \in {\mathfrak B}({\mathbb C})$. Actually $\beta (X)\in
{\mathfrak B}({\mathcal F})$ for an associated number field
${\mathcal F}(X)$. In fact, under the (normalized) Bloch regulator
${\mathfrak B}({\mathbb C})\rightarrow {\mathbb C}/{\mathbb Q}$,
the invariant $\beta (X)$ goes to $\{(2/\pi){\rm vol} (X)
+4\pi\sqrt{-1}CS(X)\}$. Let us assume that ${\mathcal F}(X)$ can
be embedded in ${\mathbb C}$ as an imaginary quadratic extension
of a totally real number field; then the Chern-Simons action,
$CS(X)$, is rational (conjecturably, $CS(X)$ is irrational if
${\mathcal F}(X)\cap \overline{{\mathcal F}(X)}\subset {\mathbb
R}$ \cite{Rosenberg1}).

Finally, taking into account the Thurston classification of all
possible three-geometries one can combined the volume and the
Chern-Simons invariants as a single complex invariant
\cite{Thurston}. By geometry or a geometric structure we mean a
pair $(X, \Gamma)$, that is a manifold X and a group $\Gamma$
acting transitively on $X$ with compact point stabilizers (we also
propose that the interior of every compact three-manifold has a
canonical decomposition into pieces which have geometric
structure). Then $\Gamma_n\backslash G_n$,\,\, $n\in (1, ..., 8)$,
represents one of the eight geometries, where $\Gamma_n$ is the
(discrete) isometry group of the corresponding geometry.
Therefore, Thurston's complex invariant can be presented in the
form $ \exp\left\{\bigcup_{\ell=1}^{\infty} \left[(2/\pi) {\rm
Vol}(\Gamma_{n\ell}\backslash G_{n\ell})+ 4\pi \sqrt{-1}
CS(\Gamma_{n\ell}\backslash G_{n\ell})\right] \right\}. $

\section{Appendix}

This appendix is devoted to a collection of basic definitions and
results on topological K-theory, following \cite{Olsen}. The
K-group of a compact manifold $X$ is the Grothendieck group
constructed out classes $[(E,F)]$ of vector bundles $E$ and $F$
over $X$, with respect to the equivalence relation:
$(E,F)\sim(E',F')$ if there exists a vector bundle $H$ on $X$ such
that $E+F'+H\cong E'+F+H$. One important elementary property of
K-groups is that, if $f,g: X\to Y$ are two homotopic maps, they
induce the same map between the corresponding K-groups.

{\it Reduced K-theory}. Taking into account that a vector bundle
over a point is just a vector space, $K({\rm pt})= {\mathbb Z}$,
we can introduce a reduced K-theory in which the topological space
consisting of a single point has trivial cohomology,
${\widetilde{K}}({\rm pt})=0$, and also ${\widetilde {K}}(X)=0$
for any contractible space $X$. Let us consider the collapsing and
inclusion maps: $ p:X\rightarrow {\rm pt}\,,\,\,\iota: {\rm
pt}\hookrightarrow X $ for a fixed base point of $X$. These maps
induce an epimorphism and a monomorphism of the corresponding
K-groups: $ p^*: K({\rm pt}) = {\mathbb Z}\rightarrow K(X)\,, $ $
\iota^*: K(X)\rightarrow K({\rm pt})= {\mathbb Z}. $ The exact
sequences of groups are:
$$
0\rightarrow {\mathbb Z} \stackrel{p^*}{\rightarrow} K(X)
\rightarrow {\widetilde{K}}(X)
\rightarrow 0\,,\,\,\,\,\,\,\,
0 \rightarrow {\widetilde {K}}(X)\rightarrow K(X)
\stackrel{\iota^*}{\rightarrow} {\mathbb Z}\,.
$$
The kernel of the map $i^*$ (or the cokernel of the map $p^*$) is
called the {\it reduced K-theory group} and is denoted by
${\widetilde {K}}(X)$, $ {\widetilde {{K}}}(X) = {\rm ker} \iota^*
= {\rm coker}\,p^*. $ Therefore, there is the fundamental
decomposition $ K(X)= {\mathbb Z}\oplus \widetilde{{K}}(X)\ . $
When $X$ is not compact, we can define $K_c(X)$, the K-theory with
compact support. It is isomorphic to ${\widetilde {{K}}}(X)$.

{\it Higher K-groups and Bott periodicity}. The {\it higher}
K-groups are labelled by a positive integer $\ell = {\mathbb Z}_+$
and can be defined according to $ K^{-\ell}(X)= K(\Sigma^\ell X)\
, $ where $\Sigma^\ell X\equiv{\mathbb S}^\ell\wedge X$ is the
$\ell-$th reduced suspension of the topological space $X$.  In
addition, $X\wedge Y= X\times Y/ (X\vee Y)$ is the smash product
of $X$ and $Y$, where $X\vee Y$ is the reduced join (the disjoint union
with a base point in each space identified). The higher K-groups
(with compact support $K_c$) can also be defined through the
suspension isomorphism: $ K^{-\ell}(X)=K(X\times {\mathbb
R}^\ell)\,. $

The {\it Bott periodicity theorem}, states that the complex
K-theory functor $K^{-\ell}$ is periodic with period two:
\begin{equation}
K^{-\ell}(X)=K^{-\ell-2}(X)
\mbox{.}
\end{equation}
The same statement holds for the reduced functor
$\widetilde{K}^{-\ell}$. This periodicity property is at variance
with the conventional cohomology theories. Note however, that the
higher reduced and unreduced K-groups differ according to $
K^{-\ell}(X)=\widetilde{K}^{-\ell}(X)\oplus K^{-\ell}({\rm pt})\ .
$
It follows that for the decomposition $X=X_1\amalg
X_2\amalg\cdots\amalg X_\ell$ of $X$ into a disjoint union of open
subspaces, the inclusions of the $X_j$ into $X$ induce a
decomposition of K-groups as $K^{-\ell}(X)=K^{-\ell}(X_1)\oplus
K^{-\ell}(X_2)\oplus \cdots\oplus K^{-\ell}(X_\ell)$. This
statement is not true for the reduced K-functor.

{\it Multiplicative structures}.
Since $K(X)$ and $\widetilde{K}(X)$ are rings (as in any cohomology
theory), the multiplication is induced by the tensor product
$E\otimes F$ of vector bundles over $X\times X$ (see for detail
\cite{Karoubi,Olsen}):
\begin{equation}
K(X)\otimes_{\mathbb Z} K(X)\rightarrow K(X)
\mbox{.}
\end{equation}
There is a homomorphism called the external tensor product or
{\it cup product},
\begin{equation}
K(X)\otimes_{\mathbb Z} K(Y)\rightarrow K(X\times Y)\,
\mbox{.}
\end{equation}
It is defined as follows. Consider the canonical projections $
\pi_{X}:X\times Y\rightarrow X $ and $ \pi_Y:X\times Y\rightarrow
Y, $ which induce homomorphisms between K-groups: $ \pi_X^{*}:
K(X)\rightarrow K(X\times Y)\,, \pi_Y^{*}: K(Y)\rightarrow
K(X\times Y). $ The cup product of $([E],[F])\in
K(X)\otimes_{\mathbb Z}K(Y)$ is the class $[E]\otimes [F]$ in
$K(X\times Y)$, where $ [E]\otimes [F]\equiv\pi_X^{*}([E])\otimes
\pi_Y^{*}([F]). $ If we consider the canonical injective inclusion
and surjective projection maps, $ X\vee Y \hookrightarrow X\times
Y \rightarrow X\wedge Y\,, $ then the contravariant functor
$\widetilde{K}^{-\ell}$ induces a split short exact sequence of
K-groups:
$$
0\longrightarrow \widetilde{K}^{-n}(X\wedge Y) \longrightarrow
\widetilde{K}^{-n}(X\times Y)\longrightarrow
\widetilde{K}^{-n}(X\vee Y) \longrightarrow 0\,.
$$
Then a useful formula for computing the K-groups of Cartesian
products \cite{Olsen} follows:
\begin{eqnarray}
\widetilde{K}^{-\ell}(X\times Y) & = & \widetilde{K}^{-\ell}
(X\wedge Y)\oplus \widetilde{K}^{-\ell}(X\vee Y)
\nonumber\\
&=& \widetilde{K}^{-\ell}(X\wedge Y)\oplus \widetilde{K}^{-\ell}(X)
\oplus\widetilde{K}^{-\ell}(Y)
\mbox{.}
\label{Kiso}
\end{eqnarray}
As an example, let $Y={\mathbb S}^1$. $K^{-1}(X)$ can be
identified with the set of K-theory classes in
$\widetilde{K}(X\times {\mathbb S}^1)$ which vanish when
restricted to $X\times {\rm pt}$. Since ${\widetilde K}({\mathbb
S}^1) =0,\, {\widetilde K}({\mathbb S}^1\wedge X) =K^{-1}(X)$, we
get
\begin{eqnarray}
\widetilde{K}(X\times {\mathbb S}^1) & = &
\widetilde{K}(X\wedge
{\mathbb S}^1)\oplus \widetilde{K}(X)\oplus
\widetilde{K}({\mathbb S}^1) = K^{-1}(X)\oplus \widetilde{K}(X)
\label{kmxs1}
\,,
\\
K^{-1}(X\times {\mathbb S}^1) & = &
K^{-1}(X\wedge {\mathbb S}^1)\oplus K^{-1}(X)\oplus K^{-1}({\mathbb S}^1)
\nonumber\\
& = & \widetilde{K}(X)\oplus K^{-1}(X)\oplus {\mathbb Z}\,
\mbox{.}
\label{kmxs}
\end{eqnarray}
Using the action of cup product on reduced K-theory we can also obtain
the following formula: $ (\widetilde{K}(X)\otimes_{\mathbb
Z}\widetilde{K}(Y))\oplus {\Xi}\rightarrow \widetilde{K}(X\wedge
Y)\oplus {\Xi}, $ here
${\Xi}=\widetilde{K}(X)\oplus\widetilde{K}(Y)\oplus{\mathbb Z}$.
The group ${\Xi}$ appears on both sides of the previous formula, and
we can eliminate it (by an appropriate restriction) and get the
homomorphism $ \widetilde{K}(X)\otimes_{\mathbb
Z}\widetilde{K}(Y)\rightarrow \widetilde{K}(X\wedge Y). $ If
$K(X)$ or $K(Y)$ is a free abelian group, this mapping and the cup
product are isomorphisms. Define $K^\#(X)$ to be the ${\mathbb
Z}_2-$graded ring $K^\#(X)=K(X)\oplus K^{-1}(X)$. If $K^\#(X)$ or
$K^\#(Y)$ is freely generated, the K-theory analog of the
cohomological K\"unneth theorem holds: $ K^\#(X\times
Y)=K^\#(X)\otimes_{\mathbb Z} K^\#(Y). $ In general, however, there
are correction terms on the right-hand side of this formula, which
relate to the torsion subgroups of the K-groups \cite{Atiyah245}.
The case of torsion-free subgroups also leads to
\begin{eqnarray}
K(X\times Y) & = & (K(X)\otimes_{\mathbb Z} K(Y))
\oplus (K^{-1}(X)\otimes_{\mathbb Z}K^{-1}(Y))\ ,
\label{KXY1cup}
\\
K^{-1}(X\times Y)& = & (K(X)\otimes_{\mathbb Z}K^{-1}(Y))
\oplus (K^{-1}(X)\otimes_{\mathbb Z} K(Y))\,.
\label{KXYcup}
\end{eqnarray}

{\it Relative K-groups}. Let us suppose that $Y$ is a closed
submanifold of $X$, then the K-group $K(X,Y)$ can be defined,
whose classes are identified with pairs of bundles over $X/Y$. If
$Y\neq \emptyset$, then the topological coset $X/Y$ is defined to
be the space $X$ with $Y$ shrunk to a point; if $Y$ is empty then
we can identify $X/Y$ with the one-point compactification $X^+$ of
$X$. We can construct a one-to-one correspondence between vector
bundles over the quotient space $X/Y$ and vector bundles $E$ over
$X$ whose restriction to $Y$ is a trivial bundle. Then the {\it
relative K-group} is defined as $ K(X,Y)\equiv \widetilde{K}(X/Y).
$ $K(X,Y)$ is a contravariant functor of the pair $(X,Y)$ and,
since $K(X)=\widetilde {K}(X)\oplus K(pt)=\widetilde{K}(X^+)$, we
have $K(X,\emptyset)= K(X)$. For the relative K-groups the Bott
periodicity holds: $ K^{-\ell}(X,Y)= K^{-\ell-2}(X,Y). $ One of
the most important properties of K-groups is that they satisfy
the Barratt-Puppe exact sequence:
\begin{equation}
\ldots\longrightarrow K^{-\ell-1}(X)\longrightarrow
K^{-\ell-1}(Y)\stackrel{\partial^*}{\longrightarrow}
K^{-\ell}(X,Y)\longrightarrow\nonumber K^{-\ell}(X)
\longrightarrow K^{-\ell}(Y)
\stackrel{\partial^*}{\longrightarrow}\ldots
\mbox{,}
\label{poppe}
\end{equation}
where $\partial$ is the boundary homomorphism. The sequence
(\ref{poppe}) connects the K-groups of $X$ and $Y\subset X$ and
makes K-theory similar to a cohomology theory. Let
$\iota:Y\rightarrow X$ and $j:(X,\emptyset)\rightarrow (X,Y)$ be
inclusions. Then, $ K(X,Y)\stackrel{j^*}{\rightarrow}
K(X)\stackrel{\iota^*} {\rightarrow}K(Y). $ If we suppose now that
$Y$ is a retract of $X$, it means that the inclusion map
$\iota:Y\rightarrow X$ admits a left inverse, then the last
sequence splits, giving
\begin{equation}
K^{-\ell}(X)= K^{-\ell}(X,Y)\oplus K^{-\ell}(Y)\,.
\label{KD}
\end{equation}

{\it Equivariant K-theory}. The natural K-theoretic methods for
classification of D-branes in orbifolds (and orientifolds) of the
Type II and Type I theories is equivariant K-theory. Let $X$ be a
smooth manifold and $G$ a group acting on $X$ ($G$ is either a
finite group or a compact Lie group). Thus, $X$ is a {\it
G-manifold} and we can write the G-action $G\times X\rightarrow X$
as $(g,x)\mapsto g\cdot x$. Suppose that a {\it G-map}
$f:X\rightarrow Y$ between two G-manifolds is a smooth map which
commutes with the action of $G$ on $X$ and $Y$: $ f(g\cdot
x)=g\cdot f(x)\,. $ Then we say that $f$ is G-equivariant. A
principal fiber bundle $E\rightarrow X$ is a {\it G-bundle} when
$E$ is a G-manifold and the canonical fiber projection $\pi$ is a
G-map: $\pi(g\cdot v)=g\cdot \pi(v)$, $\forall v\in E,\ g\in G$. A
{\it G-isomorphism} is a map $E_G\rightarrow F_G$ between
G-bundles over $X$, which is both a bundle isomorphism and a
G-map. It defines the category \footnote{ A {\it category}
${\mathcal C}$ consists the following data: 1) A class Ob
${\mathcal C}$ of objects $A, B, C, ...$; 2) A family of disjoint
sets of morphisms ${\rm Hom}(A, B)$, one for each ordered pair $A,
B$ of objects; 3) A family of maps ${\rm Hom}(A, B)\times {\rm
Hom}(B, C)\rightarrow {\rm Hom}(A, C)$, one for each ordered
triplet $A, B, C$ of objects. These data obey the axioms: a) If
$f: A\rightarrow B,\, g: B\rightarrow C,\, h: C\rightarrow D$,
then composition of morphisms is associative, that is,
$h(gf)=(hg)f$; b) To each object $B$ there exists a morphism $1_B:
B\rightarrow B$ such that $1_Bf = f,\, g1_B = g$ for $f:
A\rightarrow B$ and $g: B\rightarrow C$.} of G-equivariant bundles
over the G-space $X$. The Grothendieck group of this theory is
called G-equivariant and denoted $K_{G}(X)$. $K_{G}(X)$ consists
of pairs of bundles $(E,F)$ with G-action, modulo the equivalence
relation $(E,F)\sim (E\oplus H, F\oplus H)$ for any G-bundle $H$
over $X$. In this way D-brane configurations on $X/G$ are
understood as G-invariant configurations of D-branes on $X$,
\cite{Douglas}, in other words, the orbifold spacetime is regarded
as a G-space. For type IIB superstrings on an orbifold $X/G$, the
D-brane charge takes values in $K_G(X)$.  For type IIA one has
$K^{-1}_{G}(X)$ and for type I we have $KO_{G}(X)$ (see below). We get:
$K_G^{-1}(X)\equiv K_{G}(\Sigma X) =K_{G}({\mathbb S}^1\wedge X)$
with $G$ acting trivially on ${\mathbb S}^1$. For trivial action
of $G$ on $X$ one gets, \cite{Olsen}, $ K_G(X)= K(X)\otimes {\rm
R}(G)\,, $ where $K(X)$ is the ordinary K-group of $X$ and ${\rm R}(G)$
is the representation ring og $G$. For any
compact G-space $X$, the collapsing map $X\rightarrow$ pt gives
rise to an ${\rm R}(G)-$module structure on $K^\#_{G}(X)$, such
that ${\rm R}(G)$ is the coefficient ring in the equivariant
K-theory (instead of ${\mathbb Z}$ as in the ordinary case).
$K_{G}$ is functorial with respect to group homomorphisms. Since
$K_{G}(X)$ is a generalization of the two important classification
groups $K(X)$ and ${\rm R}(G)$, the equivariant K-theory unifies
K-theory and group representation theory. For the trivial space
$X={\rm pt}$ one has $K_G(X)={\rm R}(G)$ while the trivial group
$G={\rm Id}$ leads to $K_G(X)=K(X)$. If $H$ is a closed subgroup
of $G$, then for any H-space $X$, the inclusion
$\iota:H\hookrightarrow G$ induces an isomorphism $\iota^*:
K_{G}(G\times_HX)\stackrel{\approx}{\rightarrow}K_H(X)$.

Let the group $G$ act freely on $X$. For instance, let $G =
\Gamma$ be a co-compact group acting on real hyperbolic spaces $X
= {\mathbb H}^N$ without fixed points, giving rise to compact spaces
$X_{\Gamma}$. $X_{\Gamma}$ is a topological space and its
$\Gamma-$equivariant K-theory is just $K_{\Gamma}({\mathbb H}^N)=
K(\Gamma\backslash {\mathbb H}^N)$.

In general ($X/G$ is not a topological space) there is a useful
exact sequence for computing equivariant K-theory:
\begin{eqnarray}
K_{G}^{-1}(X,Y) & \longrightarrow & K_{G}^{-1}(X) \longrightarrow
K_{G}^{-1}(Y) \stackrel{\partial^*}{\longrightarrow} K_{G}(X,Y)
\nonumber \\
K_{G}^{-1}(X,Y) & \stackrel{\partial^*}\longleftarrow & K_G(Y)
\longleftarrow K_G(Y) \longrightarrow
K_G(X)\longleftarrow K_G(X,Y)
\label{sixterm}
\end{eqnarray}
Here $Y$ is a closed G-subspace of a locally compact G-space $X$,
and the relative K-theory is defined by
$K_{G}^{-\ell}(X,Y)=\widetilde{K}_{G}^{-\ell}(X/Y)$.

{\it Real K-groups}. Let us consider pairs of bundles $(E,F)$ with
$(E\oplus H,F\oplus H)$\, for any $SO(K)$ bundle $H$. Pairs
$(E,F)$ with this equivalence relation define the real $K-$group
$KO(X)$ of the spacetime $X$. Alternatively we can introduce the
reduced real K-theory group $\widetilde{KO}(X)$. It follows from
the bound state construction that D-brane configurations of Type I
superstring theory are classified by $KO(X)$ with compact support
\cite{Witten}. The torsion KO-groups modify various product
relations that have been considered above. Using, for example,
(\ref{Kiso}) we get the analog of (\ref{kmxs1}):
\begin{equation}
\widetilde{KO}(X\times {\mathbb S}^1)=\widetilde{KO}^{-1}(X)\oplus
\widetilde{KO}(X)\oplus{\mathbb Z}_2\,.
\label{KOXS1}
\end{equation}
Remember that the Grothendieck group of all virtual bundles with
involutions on $X$ is called the real K-group $KR(X)$. In the
standard way one can define higher groups $KR^{-m}(X)$ by $
\widetilde{KR}^{-\ell}(X)=\widetilde{KR}(X\wedge {\mathbb
S}^\ell)\,, $ with the involution $\tau$ on $X$, i.e. a
homeomorphism $\tau: X\rightarrow X$, $\tau^2 =Id_X$, extended to
$X\wedge {\mathbb S}^\ell$ by a trivial action on ${\mathbb
S}^\ell$. Let ${\mathbb Z}^{p,q}$ be the $(p+q)-$dimensional real
space in which an involution acts as a reflection of the last $q$
coordinates: for given $(x,y)\in {\mathbb R}^p\times{\mathbb R}^q$
we get $\tau:(x,y)\mapsto (x,-y)$. Suppose ${\mathbb S}^{p,q}$ is
the unit sphere of dimension $p+q-1$ in ${\mathbb R}^{p,q}$ with
respect to the flat Euclidean metric on ${\mathbb
R}^p\times{\mathbb R}^q$. One can define a two-parameter set of
higher degree KR-groups \cite{Olsen}: $
\widetilde{KR}^{p,q}(X)=\widetilde{KR}(X\wedge{\mathbb
R}_+^{p,q})\, $ and $ KR^{p,q}(X)=KR(X\times {\mathbb R}^{p,q})\,.
$ Then we have $ KR^{-\ell}(X)= KR^{\ell,0}(X)\,, $ and Bott
periodicity in KR-theory has the form
\begin{equation}
KR^{p,q}(X) = KR^{p+1,q+1}(X)\,,\,\,\,\,\,
KR^{-\ell}(X) = KR^{-\ell-8}(X)
\mbox{.}
\end{equation}
>From this formula we get: $KR^{p,q}(X)= KR^{q-p}(X)$, and
$KR^{p,q}(X)$ only depends on the difference $p-q$. Also,
$KR^{p,q}(X)$ depends only on this difference modulo 8. In fact,
we can define negative-dimensional spheres as those with antipodal
involutions in KR-theory, with ${\mathbb S}^{\ell,0}$ being
identified as ${\mathbb S}^{\ell-1}$ and ${\mathbb S}^{0,\ell}$ as
${\mathbb S}^{-\ell-1}$. Identifying ${\mathbb R}^{1,1}={\mathbb
C}$ with the involution $\tau$ acting as the complex conjugation,
one gets the $(1,1)$ periodicity theorem in the following form $
KR(X)=KR(X\times {\mathbb C})\,, $ which holds for any locally
compact space $X$. For trivial $\tau-$action on $X$ we have
\begin{equation}
KR^{-\ell}(X\times {\mathbb S}^{0,1}) = K^{-\ell}(X)\,,
\,\,\,\,\,\,\,\,
KR^{-\ell}(X) = KO^{-\ell}(X)\ .
\label{KRKO}
\end{equation}
Most of the properties discussed above for K-groups have obvious
counterparts in the real case. For example, by repeating the
calculations which give (\ref{kmxs}) we can obtain, for a trivial
action of $\tau$ on $X$, the product formula \cite{Olsen}:
\begin{eqnarray}
\!\!\!\!\!\!\!\!\!\!\!\!\!\!
\widetilde{KR}^{-1}(X\times{\mathbb S}^{1,1})&=&\widetilde{KR}^{-1}
(X\wedge{\mathbb S}^{1,1})\oplus\widetilde{KR}^{-1}(X)\oplus
\widetilde{KR}^{-1}({\mathbb S}^{1,1})
\nonumber\\
&=&\widetilde{KR}^{1,1}(X)\oplus\widetilde{KO}^{-1}(X)
\oplus{\mathbb Z}
=\widetilde{KO}(X)\oplus\widetilde{KO}^{-1}(X)\oplus{\mathbb Z}
\mbox{.}
\label{KRS11}
\end{eqnarray}


\begin{thebibliography}{10}



\bibitem{Witten} E.~Witten, {\it D-branes and K-theory},
JHEP {\bf 9812} (1998) 019
[hep-th/9810188].

\bibitem{Gukov}
S. Gukov, {\it K-Theory, Reality, and Orientifolds}, Commun.
Math. Phys. {\bf 210} (2000) 621 [hep-th/9901042].

\bibitem{Sharpe}
E. Sharpe, {\it D-Branes, Derived Categories, and Grothendieck
Groups}, Nucl. Phys. {\bf B 561} (1999) 433 [hep-th/9902116].

\bibitem{Atiyah}
M. F. Atiyah, {\it K-Theory}, Benjamin, New York, 1967.

\bibitem{Karoubi}
M. Karoubi, {\it K-Theory. An Introduction}, Springer-Verlag,
Berlin, 1978.

\bibitem{Husemoller}
D. Husemoller, {\it Fibre Bundles}. McGraw-Hill, New York, 1966.

\bibitem{Diaconescu}
D.~Diaconescu and J.~Gomis,
{\it Fractional branes and boundary states in orbifold theories},
JHEP {\bf 0010} (2000) 001 [hep-th/9906242].

\bibitem{Garcia}
H.~Garcia-Compean, {\it D-branes in orbifold singularities and
equivariant K-theory}, Nucl. Phys. {\bf B 557} (1999) 480
[hep-th/9812226].

\bibitem{Bergman}
O.~Bergman, E.~Gimon, and B.~Kol,
{\it Strings on Orbifold Lines}, JHEP {\bf 0105} (2001) 019
[hep-th/0102095].

\bibitem{Ferrara}
S. Ferrara, A Kehagias, H. Partouche and A. Zaffaroni, {\it
Membranes and Fivebranes with Lower Supersymmetry and their AdS
Supergravity Duals}, Phys. Lett. {\bf B 431} (1998) 42
[hep-th/9803109].

\bibitem{Bytsenko22}
A. A. Bytsenko, M. E. X. Guimar\~aes and J. A. Helayel-Neto,
{\it Hyperbolic Space Forms and Orbifold Compactification in M-Theory},
Proc. Sci. WC2004 (2004) 17.

\bibitem{BytsenkoM}
A. A. Bytsenko, M. E. X. Guimar\~aes and R. Kerner, {\it Orbifold
Compactification and Solutions of M-Theory from Milne Spaces},
Eur. Phys. J. {\bf C 39} (2005) 519.

\bibitem{Olsen}
K. Olsen and R. J. Szabo, {\it Constructing D-Branes From
K-Theory}, Adv. Theor. Math. Phys. {\bf 3} (1999) 889
[hep-th/9907140].

\bibitem{Moore}
R. Minasian and G. Moore, {\it K-theory and Ramond-Ramond
Charges}, JHEP {\bf 9711} (1997) 002 [hep-th/9710230].

\bibitem{Moore00}
G. Moore and E. Witten, {\it Self-duality, Ramond-Ramond fields, and
K-theory}, JHEP {\bf 0005} (2000) 032 [hep-th/9912279].

\bibitem{Switzer}
R. M. Switzer, {\it Algebraic Topology: An Introduction},
Springer-Verlag, 1978.

\bibitem{Reis}
R. M. G. Reis and R. J. Szabo, {\it Geometric K-Homology of Flat
D-Branes}, hep-th/0507043.

\bibitem{Cheung} Y. K. Cheung and Z. Yin, {\it Anomalies, branes and currents},
Nucl. Phys. {\bf B 517} (1998) 69 [hep-th/9710206].

\bibitem{Atiyah62}
M. F. Atiyah and F. Hirzebruch, {\it Analytical Cycles on Complex
Manifolds}, Topology {\bf 1} (1962) 25.

\bibitem{Boer}
J. de Boer, R. Dijgraaf, K. Hori, A. Keurentjes, J. Morgan, D. R.
Morrison and S. Sethi, {\it Triples, Fluxes, and Strings}, Adv.
Teor. Math. Phys. {\bf 4} (2002) 995 [hep-th/0103170].

\bibitem{Moscovici1}
H. Moscovici and R. Stanton, {\it Eta invariants of Dirac
operators on locally symmetric manifolds}, Invent. Math. {\bf 95}
(1989) 629.

\bibitem{Hotta}
R. Hotta and R. Parthasarathy, {\it A geometric meaning of the
multiplicity of integrable discrete classes in
$L^2(\Gamma\backslash G)$}, Osaka J. Math. {\bf 10} (1973) 211.

\bibitem{Atiyah11}
M. Atiyah and I. M. Singer, {\it The index of elliptic operators:
III}, Ann. of Math. {\bf 87} (1968) 546.

\bibitem{Atiyah1}
M. F. Atiyah, V. K. Patodi and I. M. Singer, {\it Spectral
Asymmetry and Riemannian Geometry. I}, Math. Proc. Camb. Phil.
Soc. {\bf 77} (1975) 43.

\bibitem{Atiyah2}
M. F. Atiyah, V. K. Patodi and I. M. Singer, {\it Spectral
Asymmetry and Riemannian Geometry. II}, Math. Proc. Camb. Phil.
Soc. {\bf 78} (1975) 405.

\bibitem{Atiyah3}
M. F. Atiyah, V. K. Patodi and I. M. Singer, {\it Spectral
Asymmetry and Riemannian Geometry. III}, Math. Proc. Camb. Phil.
Soc. {\bf 79} (1976) 71.

\bibitem{Laws89}
H. B. Lawson and M. L. Michelson, {\it Spin Geometry}, Princeton
Mathematical Series {\bf 38}, Princeton Univ. Press, 1989.

\bibitem{Barbasch}
D. Barbasch and H. Moscovici, {\it $L^2-$index and the Selberg
trace formula}, J. Funct. Anal. {\bf 53} (1983) 151.

\bibitem{WO}
N. E. Wegge-Olsen, {\it K-theory and $C^*-$algebras}, Oxford
University Press, New York, 1993.

\bibitem{Higson}
N. Higson and J. Roe, {\it Analytic K-Homology}, Oxford
Mathematical Monographs, Oxford University Press, 2000.

\bibitem{Atiyah64}
M. F. Atiyah, R. Bott, and A. Shapiro, {\it Clifford modules},
Topology {\bf 3} (1964), no. Suppl. 1, 3.

\bibitem{Brown}
L. Brown, R. Douglas and P. Fillmore, Ann.
of Math. {\bf 105} (1977) 265.

\bibitem{Kasparov1}
G. G. Kasparov, {\it Equivariant KK-theory and the Novikov
conjecture}, Invent. Math. {\bf 91} (1988) 147.

\bibitem{Connes94}
A. Connes, {\it Noncommutative geometry}, Academic Press, 1994.

\bibitem{Carey}
A. L. Carey, K. C. Hannabuss, V. Mathai and P. McCann, {\it
Quantum Hall Effect on the Hyperbolic Plane}, Commun. Math. Phys.
{\bf 190} (1998) 629.

\bibitem{Kasparov95}
G. Kasparov, {\it K-theory, group $C^*-$algebras and higher
signatures}, Conspectus, 1980, published in {\it Novikov
conjectures, index theorems and rigidity}, vol. {\bf 1}, Editors
S. Ferry, A. Ranicki and J. Rosenberg, Lond. Math. Soc. Lecture
Note Series {\bf 226}, Cambridge University Press, 1995.

\bibitem{Kasparov}
G. Kasparov, {\it Lorentz groups, K-theory of unitary
representations and crossed products}, Soviet. Math. Dokl. {\bf
29} (1984) 256.

\bibitem{Julg}
P. Julg and G. Kasparov, {\it Operator K-theory for the group
$SU(n,1)$}, J. Reine Angew. Math. {\bf 463} (1995) 99.

\bibitem{Bouwknegt}
P. Bouwknegt and V. Mathai, {\it D-Branes, B-Fields and twisted
K-theory}, JHEP {\bf 007} (2000) [hep-th/0002023].

\bibitem{Dixmier}
J. Dixmier and A. Douady, {\it Champs continues d'espaces
hilbertiens at de $C^{*}-$algebres}, Bull. Soc. Math. France {\bf
91} (1963) 227.

\bibitem{Rosenberg}
J. Rosenberg, {\it Continuous trace algebras from the bundle
theoretic point of view}, Jour. Aus. Math. Soc. {\bf 47} (1989)
368.

\bibitem{Elliott}
G. Elliott, {\it On the K-theory of the $C^*-$algebra generated by
a projective representation of atorsion-free discrete group}, In:
{\it Operator Algebras and Group Representations}, London, Pitman
(1983) 157.

\bibitem{Connes}
A. Connes, {Noncommutative differential geometry}, Publ. Math.
I.H.E.S. {\bf 62} (1986) 257.

\bibitem{Packer1}
J. Packer and I. Raeburn, {\it Twisted cross products of
$C^*-$algebras}, Math. Proc. Camb. Phil. Soc. {\bf 106} (1989)
293.

\bibitem{Suslin1}
A. Suslin, {\it On the K-theory of local fields}, J. Pure Appl.
Algebra {\bf 34} (1984) 301.

\bibitem{Suslin3}
A. A. Suslin, {\it $K_3$ of a field, and the Bloch group, Galois
theory, rings, algebraic groups and their applications}, Trudy
Mat. Inst. Steklov. {\bf 183} (1990) 180, 229.

\bibitem{Suslin87}
A. A. Suslin, {\it Algebraic K-theory of fields}, Proc. Int. Cong.
Math. Berkeley 1986, {\bf 1} (1987) 222.

\bibitem{Neumann1}
W. D. Neumann and J. Yang, {\it Rationality problems for K-theory
and Chern-Simons invariants of hyperbolic 3-manifolds}, Ens. Mat.
{\bf 41} (1995) 281.

\bibitem{Rosenberg1}
J. Rosenberg, {\it Recent Progress in Algebraic $K-$Theory and its
Relationship with Toplogy and Analysis}, Mini-Course for the Joint
Summer Research Conference on Algebraic K-Theory, Seatle, July,
1997.

\bibitem{Thurston}
W. Thurston, {\it Three-Dimensional Manifolds, Kleinian Groups and
Hyperbolic Geometry}, Bull. Amer. Math. Soc. (N.S.) {\bf 6} (1982)
357.

\bibitem{Atiyah245}
M. F. Atiyah, {\it Vector Bundles and the K\"unneth Formula},
Topology {\bf 1} (1962) 245.

\bibitem{Douglas}
M. R. Douglas and G. Moore, {\it D-branes, Quivers and ALE
Instantons}, hep-th/9603167.



\end{thebibliography}
\end{document}